\documentclass[12pt,preprint]{aastex}
\slugcomment{Accepted and scheduled for publication  
 {\it The Astrophysical Journal},  for the ApJ July 1, 2014, v 789, 1 issue}  
\def\lax {\ifmmode{_<\atop^{\sim}}\else{${_<\atop^{\sim}}$}\fi}  
\def\gax {\ifmmode{_>\atop^{\sim}}\else{${_>\atop^{\sim}}$}\fi}  
\def\gtorder{\mathrel{\raise.3ex\hbox{$>$}\mkern-14mu
             \lower0.6ex\hbox{$\sim$}}}

\def\cm2{cm$^{-2}$}
\def\s1{s$^{-1}$}

\begin{document}

\title{Black hole mass determination in the X-ray binary 4U~1630--47:  Scaling of 
spectral and variability characteristics}






\author{Elena Seifina\altaffilmark{1}, Lev Titarchuk\altaffilmark{2} and 
Nikolai Shaposhnikov\altaffilmark{3}}

\altaffiltext{1}{Moscow M.V.~Lomonosov State University/Sternberg Astronomical Institute, Universitetsky 
Prospect 13, Moscow, 119992, Russia; seif@sai.msu.ru}
\altaffiltext{2}{Dipartimento di Fisica, Universit\`a di Ferrara, Via Saragat 1, I-44122 Ferrara, Italy, email:titarchuk@fe.infn.it; George Mason University Fairfax, VA 22030;   
Goddard Space Flight Center, NASA,  code 663, Greenbelt  
MD 20771, USA; email:lev@milkyway.gsfc.nasa.gov}
\altaffiltext{3}{CRESST/University of Maryland, Department of Astronomy, College Park MD 20742, 
Goddard Space Flight Center, NASA,  code 663, Greenbelt  
MD 20771,  USA: email:nikolai.v.shaposhnikov@nasa.gov USA
}

\begin{abstract}
We present the results of a comprehensive investigation on the evolution of spectral and timing properties of  
the Galactic black hole candidate 4U~1630--47 during its spectral transitions. 
 In particular, we  show how  a  scaling  of the  correlation of 
the photon index of the Comptonized spectral component  $\Gamma$ with 
low frequency of quasi-periodic  oscillations (QPO), $\nu_L$  and 
mass accretion  rate, $\dot M$ can be applied to the black hole mass  and the inclination angle estimates. 
We analyze 
the transition episodes observed with the {\it Rossi} X-ray Timing Explorer
({\it RXTE}) and {\it Beppo}SAX satellites. 
We find that the broadband X-ray energy spectra of 4U~1630--47 during
all spectral states can be modeled by a combination of a thermal  component, a Comptonized
component 
and  a red-skewed iron line component. We also
establish that  $\Gamma$ monotonically increases during  
transition from the low-hard state to the high-soft
state and then saturates  for high mass accretion rates.
The 
index saturation levels vary for  different transition episodes.
Correlations of  $\Gamma$ versus 
$\nu_L$ also shows saturation at  $\Gamma\sim 3$.    
$\Gamma-\dot M$ and $\Gamma-\nu_L$  correlations with their index saturation revealed in 4U~1630--47 are similar
to those established  in a number of  other BHCs and can be considered
as an observational evidence for the presence of a black hole in these sources.
The scaling technique, which relies on  XTE~J1550--564, GRO~1655-40 and H~1743-322
 as   reference sources, allows us to evaluate  a black hole mass   in  4U~1630--47 yielding
 $M_{BH}\sim 10\pm 0.1$ solar masses,  and to constrain
the inclination angle of $i\lax 70^{\circ}$.

\end{abstract}


\section{Introduction}

The problem of the dynamical mass determination of 
black holes  (BHs) in binary systems
is closely  related to mass function and mass ratio estimates   (using optical counterpart data) and also to the orbital inclination angle \citep[see][and references therein]{Orosz2003}. 
This  mass determination requires a knowledge of the period and radial velocity measurement. 
The
absence 
of X-ray eclipses and the rotational broadening of absorption lines (radial velocity shifts) in the optical star spectrum also  
facilitate BH mass evaluation. However, there are relatively  a few   of low-mass binary sources, for which 
optical  emission 
is  accessible, as a result of Galactic extinction. 
On the other hand,
it is possible to evaluate   the mass of the central object (for example,  a BH mass)   
 based on  X-ray data  and  timing characteristics (even when conventional dynamical methods cannot be used). Furthermore, only in transient systems where accretion shuts down and the system reverts to a quiescent state can the Keplerian  photospheric line shifts be measured. 

A new method of the BH mass determination was developed by  \cite{st09}, hereafter ST09, 
using 
correlation scaling  between X-ray spectral and timing properties
observed from many Galactic BH 
binaries during their hard-soft state transitions.
ST09 consider the {\it Transition layer} (TL) model proposed by Titarchuk, Lapidus \& Muslimov (1998), hereafter TLM98 (see also Titarchuk  \& Osherovich 1999, hereafter TO99) in which the TL size  is proportional to BH mass.
 The Comptonization parameter  $Y$, which is a product of the average number of scatterings and the efficiency of up-scattering, is inversely proportional to the energy spectral index $\alpha$ (ST09).
TLM98  suggest that such a configuration is  a result 
of an adjustment  of 
a Keplerian flow (disk) to  
a turbulent, innermost sub-Keplerian flow (TL). 
The TL formation is probably related to  the shock formation near the adjustment radius (see TLM98).
Thus, the accretion flow releases its gravitational energy in  the TL where  its temperature is regulated by this gravitational energy release,   Compton cooling and illumination of the TL by external soft photon flux from the disk (see TLM98 and  TO99). The emergent spectrum is formed as a result
of upscattering of these soft (disk) photons in the relatively hot plasma of TL.  In the low-hard state, when the flux of the disk photons is relatively weak, the TL plasma temperature is $kT_e\sim  50$ keV and  the resulting  photon index is $\Gamma \lax 1.9$. On the other hand, in the high-soft state, when the flux of the disk soft photon dominates the gravitational energy release in TL, the plasma temperature  significantly decreases to 5-10 keV  and  the photon index $\Gamma$ becomes greater than 2 as a result of disk photon cooling and General Relativity (GR) effects [see \cite{tz98} and \cite{LT99}]. 

 It is well known that different 
BHCs
show different spectral-timing correlation patterns, $\Gamma$ vs quasi-periodic oscillation   (QPO) frequency $\nu_L$ and $\Gamma$ vs mass accretion rate,  depending on  mass of the compact object, the distance from the Earth observer and the binary inclination. The correlation of $\Gamma$  versus  $\nu_L$    in black binaries was discovered by Vignarca et al. (2003)  using {\it RXTE} data in terms of a phenomenological model,  a  high-energy cutoff power law, corrected for interstellar absorption, plus a Gaussian emission line to take into account an excess at 6.4 keV.   
Recently Stiele et al. (2013), hereafter SBKM13,  confirmed  {\it the correlations between the spectral 
index and QPO frequency  in a sample of Galactic black hole candidate (BHC)  binaries 
(GX 339-4, H 1743-322, and XTE J1650-500)
using {\it RXTE} data}. For the spectral fitting   they  applied  a sum of the disk emission approximated by the $diskbb$ model (Mitsuda et al. 1984) with the $simpl$ model (Steiner et al. 2009)  for Compton scattering as well as a reflection  component.
However   Vignarca et al. and Stiele et al.  did not apply these induced correlations for BH mass determination.  
 
In the studies of 
Shaposhnikov \& Titarchuk (2007, 2009), Titarchuk \& Seifina (2009) [TS09 hereafter] and Shrader et al. (2010) 
the authors demonstrate that 
these correlation tracks 
provide a wide scaling range for a source with unknown BH mass. 
 Recently,  application of this method was also extended to a study
of  another class of X-ray sources, the ultra-luminous X-ray source or ULXs , NGC~5408~X-1 in
 Strohmayer \& Mushotzky (2009).

The observed variability and spectral properties vary in a well defined manner 
throughout different spectral states.  Furthermore   these correlations are seen in many
sources, which vary widely in luminosity. The universality 
of these correlations  suggests that the basic physical 
processes are very similar for each of these sources  
and their observational manifestations are determined by a common set of basic 
physical parameters.

It this work  we take 
an opportunity to determine the  mass of the putative BH  in  the X-ray binary 4U~1630--47, 
for which the compact oject mass has  not been evaluated 
using   any traditional method (see above).
4U~1630--47 is an X-ray transient discovered by {\it Uhuru} 
[see Jones et
al. (1976)], although its first recorded outburst was apparently detected in 1969
by {\it Vela} 5B (Priedhorsky 1986). 
4U~1630--47 is a well-studied black hole candidate [Tanaka \& Lewin (1997); 
Parmar et al. (1995, 1997); 
Kuulkers et al. (1997), and Oosterbroek et al. (1998)]. 
 4U~1630--47 is an ultra-soft X-ray transient, 
which shows  various multiple outbursts.
4U~1630--47 has also been observed by 
{\it Ginga} [Kuulkers et al. (1997)], 
{\it Beppo}SAX (Oosterbroek et al., 1998), INTEGRAL [\cite{Tomsick05}, T05 hereafter] 
and {\it Suzaku} (Kubota et al., 2007).  
The 1984 outburst, and its following 
decay, were 
studied by EXOSAT (Parmar et al. 1986). 
4U~1630--47 is a part of the  
to the group of the X-ray transients known as
X-ray novae (Tanaka \& Shibazaki 1996; Sunyaev et al. 1994). All
sources of this type 
are assumed to be recurrent. However the recurrence time scales seem to range from years to decades so in some cases only a  single event have been recorded (e.g. Chen, Shrader \& Livio, 1997 and Grindlay et al. 2014).  
A typical
recurrence time for outbursts of such sources is usually  $\sim 10-50$ years. 
In this respect, 
the fact that 4U~1630--47 and also GX 339-4 exhibit relatively frequent outbursts is  
uncommon. The source 4U~1630--47  lies in the 
direction towards the Galactic center. Observations show that the source is
heavily absorbed in soft X-rays, indicating a large source distance
($\ge 10$ kpc). No optical counterpart has yet been identified, probably due to 
the large 
extinction amount of optical ($>20$ mag) and reddening. Therefore, the dynamical evidence for the presence of a black hole in this binary system is still missing.

For the reasons mentioned above the binary orbital period of 4U~1630--47 is  
not known.
The spread in estimates of the column density 
toward 4U~1630 is quite broad. Specifically, $N_H$ values as derived from spectral fits of
X-ray data are in the range from 5$\times 10^{22}$ cm$^{-2}$ (Parmar et al. 1986, 
1997; ) to 1.5$\times 10^{23}$ cm$^{-2}$ (Kuulkers et al. 1998; T05; Oosterbroek et al. 1998; 
Cui et al. 2000; Dieters et al. 2000; Tomsick \& Kaaret 2000; Trudolyubov et al.
2001). These measurements vary with   X-ray brightness of the system and clearly 
include a contribution intrinsic to the system. 
The column density estimate derived from HI radio surveys (Kerr et al. 1986) yields a value of 2$\times 10^{22}$ cm$^{-2}$.
Some difficulties of the infrared counterpart detection of this X-ray source are related to the requirement of continuous 
and deep IR monitoring   during both outburst and quiescence (Callanan et al., 2000).
Augusteijn et al. (2001) were able to detect a variable source at K=16.1 
mag located inside the radio error circle which they identify as the infrared counterpart to the X-ray source. 
They conclude  that 4U~1630--47 is most likely a black hole candidate  X-ray binary similar to GRO~J1655-40 or 4U~1543-47, containing 
a relatively early-type secondary. It should be noted that among multiple X-ray outbursts registered from 4U~1630--47,  only the 1998 outburst was accompanied by radio jet emission (Hjellming et al., 1999) detected by NRAO VLA and Australia Telescope Compact Array (ATCA). 
Based   on its similarities to the X-ray properties of 
sources in which a black hole has been identified using the dynamical methods,  4U~1630--47 is also  suggested to be a BHC
[see e.g. Parmar et al., 
(1986)].

The emission properties of accreting black holes are generally 
classified in terms of {\it canonical} ``spectral states''  (see Remillard \& McClintock
2006; Belloni 2005; Klein-Wolt \& van der Klis 2008). 
We use 
a general BH state classification which includes five basic 
BH states: {\it quiescent}, {\it low-hard}, {\it intermediate}, {\it high-soft}, and
{\it very high states} (LHS, IS, HSS and VHS respectively). 
When a BH goes 
into outburst it starts from 
its {\it quiescent} state and enters  the 
LHS, where 
the energy spectrum is characterized 
by a hard Comptonization component and   a relatively  weak thermal component. 
This spectrum 
is  a result of Comptonization of soft  photons by electrons of the hot ambient media
[the Compton cloud, herein CC] (see, e.g., Sunyaev \& Titarchuk 1980). Throughout this work we use 
terminology where the CC is associated with a transition layer (TL).
 In general, the LHS is characterized by strong variability, 
which can been seen as a flat-top broken power-law (WRN) in its power density  spectrum (PDS), along with QPOs in the range of 0.01 -- 25 Hz.

In the  HSS, the photon spectrum is well represented by
the sum of a strong 
thermal component, which is probably originated in the inner 
accretion disk, and an extended power law component.
It is worth noting that in the HSS the PDS flat-top component is absent   and all QPOs disappear.
Overall variability in the HSS is either absent or very low and presented by a weak power law in the power spectrum.   
The intermediate state (IS) is a transitional state from 
the LHS to 
the HSS and vice versa. Note that in addition to the LHS, IS and HSS we can also observe sometimes a very high state (VHS) in which a both {\it blackbody} and non-thermal  components are present. Flat-top noise and low frequency 
QPOs may be seen in their power-density  spectra (PDS). Also, this state is generally associated  with 
with  high frequency QPOs above 100 Hz.

{

Three main types of LFQPOs, 
so called Type A, B and C were originally identified in the light curve of XTE~J1550-564 (Wijnands et al. 1999; 
Remillard et al. 2002), and  further found 
in several sources (see Casella et al. 2005 and references therein).
In the context of the state classification outlined above, it is possible to attribute these three QPO types  to different  spectral conditions (see Homan et al. 2001, Homan \& Belloni 2005, Belloni et al. 2005). The type-C QPO is associated with the (radio loud) hard intermediate state 
and the low/hard (LS) state (Soleri et al., 2008). It is  commonly seen in almost all BHCs in which  a QPO  is correlated with the count rate and this QPO is characterized by a high fractional variability. In turn, C-type QPOs  demonstrate  a clear correlation with the spectral  index  and provide an opportunity for a BH mass scaling (ST07, ST09).  Casella et al. (2005) showed  that the QPO type and  frequency in BHCs change 
systematically as a function of the inverse of the source rms fractional variation. This behavior is seen 
over different BHCs and presents fair 
similarities with the LFQPOs observed in neutron star low-mass X-ray binaries.
Thus Casella et al. (2005) suggested that C, B, and A type LFQPOs in BHCs correspond to HBOs , NBOs, 
and FBOs  (horizontal, normal  and flaring branch oscillations respectively) of high-luminosity neutron star systems of the {\it Z} class.

 Based on the analysis of {\it RXTE} data T05 concluded that 4U~1630--47 exhibits many properties which are not   completely encompassed by Remillard \& McClintock (2006) state definitions. 
In particular, T05 reported on so-called {\it flaring} state  observed on timescales of 10 -- 100 s 
during 2002 -- 2004 observations. Similar high amplitude flaring behavior was previously detected only  in BHC GRS~1915+105 (Belloni et al., 2005). However, in contrast to GRS 1915+105 case, the X-ray light curves of  4U~1630--47 do not show  
repeating patterns seen in GRS 1915+105.


 In this work we study in detail the behavior of the energy and power spectra observed by {\it RXTE} and 
 {\it Beppo}SAX from the galactic
BHC  4U~1630--47 during 1996 -- 2004 outburst activity.  Specifically, we concentrate on
the phenomenology of the photon index and the QPO frequency. 
Based on the results of our data analysis we estimate  the BH mass  in 4U~1630--47.
In \S 2 we present the list of observations used in the data analysis while 
in \S 3 we provide the details of X-ray spectral analysis.  We discuss the evolution of 
X-ray spectral and timing  properties during the state transition in \S 4-\S 6.  
We  also discuss our results and make our final  conclusions in \S 7$-$\S 8.

\section{Data Selection \label{data}}

Broad band energy spectra of 4U~1630--47 were obtained
combining data from  three {\it Beppo}SAX Narrow
Field Instruments (NFIs): the Low Energy Concentrator
Spectrometer [LECS; \citet{Parmar97}] for the 0.3 -- 4 keV range, the Medium Energy Concentrator Spectrometer
[MECS; \citet{boel97}] for the 1.8 -- 10 keV range  and the Phoswich Detection
System [PDS; \citet{fron97}] for the 15 -- 200 keV range. 
The SAXDAS data analysis package is used for the data processing. 
We performed spectral analysis for each  instrument in a corresponding 
energy range with 
 a well known  response matrix.
The LECS data have been 
renormalized to match the MECS data. Relative normalizations of the NFIs were treated 
as free parameters in the
 model fits, except for the MECS normalization that was fixed at unity.  
 The cross-calibration factor obtained in this way is found to be 
in a standard range for each
 instrument
\footnote{http://heasarc.nasa.gov/docs/sax/abc/saxabc/saxabc.html}.
Additionally, 
spectra have been rebinned  in accordance with 
the energy resolution of the instruments 
using
 rebinning template files
 in GRPPHA
to obtain better signal to noise ratio. 
Systematic uncertainties  of 1\% 
have been applied to all
 spectra. 
In Table 1 we listed  the {\it Beppo}SAX observations used in the present 
analysis. 

We  also analyzed 425 {\it RXTE} observations  taken between February 1996 and May 2004.
Standard tasks of the LHEASOFT/FTOOLS
5.3 software package were 
used for data processing.
For spectral analysis we 
used PCA {\it Standard 2} mode data, collected 
in the 3 -- 23~keV energy range, applying 
PCA response 
calibration (pcarmf v11.7).
{
The fitting was carried out using the standard XSPEC v 12.6.0 fitting package \footnote{http://heasarc.gsfc.nasa.gov/FTP/sax/cal/responses/grouping}. 
The standard dead time correction procedure 
has been applied to the data. 
In order to construct broad-band spectra the data from HEXTE detectors have  been also utilized.
We  subtracted a background corrected  in  off-source observations. 
 The data are available through the GSFC public archive 
(http://heasarc.gsfc.nasa.gov). Systematic error of 0.5\% has been applied to all analyzed {\it RXTE}  spectra. 
In  Table 2 we listed the  groups
 of {\it RXTE} observations tracing 
thorough 
the source evolution during different states. 

We have  performed an analysis of {\it RXTE} observations  of 4U~1630--47  spanning eight years 
made at seven 
intervals indicated by  blue rectangles in Figure~\ref{asm_1630} ($top$).
The {\it RXTE} 
energy spectra were modeled using XSPEC astrophysical fitting software. 
 We have also used public 4U~1630--47 data from the  All-Sky Monitor (ASM) 
on-board \textit{RXTE}, which demonstrate 
long-term 
variability of the 2$-$12 keV flux 
during all observation scans. 

According to ASM monitoring 4U~1630--47 shows complex long-term variations. 
Specifically, during eight years (1996 -- 2004) the {\it RXTE} have detected five outbursts from
4U~1630--47 (see Fig.~\ref{asm_1630}). 
These outbursts follow 
the recurrence interval of about 600 days
(Parmar et al. 1995) and differ significantly in  shape. 

Possible resemblance between 
individual outbursts from 4U~1630--47 and from other known BHCs should  be emphasized.
 For example,  the outbursts of 1996 and 1999 resemble  the state transitions  
in the persistent BHC
 Cyg~X-1 [\cite{Cui98}] 
and  GX~339-4 [\cite{Belloni99}], 
while the 1998 outburst showed a fast rise and exponential decay (FRED) 
profile  which is  typical for many X-ray novae [\cite{Chen97}]. 

Data from the PCA and HEXTE detectors as well as {\it Beppo}SAX detectors have been used 
to constrain spectral fits, 
while ASM data provided long-term intensity state monitoring. Results of our long-term study 
of 4U~1630--47 are  presented in detail in the next sections and compared with 
results for XTE~J1550--564
in order to estimate  a BH mass  in  4U~1630--47.

In general, the broadband spectral sensitivities of two X-ray orbital observatories, 
{\it RXTE}  (Bradt et al., 1993) and {\it Beppo}SAX (Boella et al. 1997), combined with the high timing 
resolution of {\it RXTE}   provide 
a means to  study both the detailed broadband spectra and long-term 
spectral and timing evolution of BH hosting X-ray binaries. 


\section{Spectral Analysis \label{spectral analysis}}

Our 
spectral model is based on the following physical paradigm of the 
accretion process.
We assume that accretion onto a black hole takes place when the material passing through 
 three main  regions:  a geometrically thin accretion disk [standard Shakura-Sunyaev 
disk, see \cite{ss73}],
 the  transition layer (TL) \citep{tlm98} and the converging flow  
  \citep{tz98}. The disk  photons 
are upscattered off energetic electrons   of   the  transition layer  and the converging flow  
(see also Fig.~\ref{geometry}).
Some fraction of these seed  photons can be also seen directly by the Earth observer. 
{\it Pink} 
and {\it blue} waved arrows 
shown in Fig.~\ref{geometry}
correspond to $Comptb$ 
and blackbody 
components, respectively. 


According to the physical picture described above,  for our spectral analysis we 
use a model which consists of  sum of a  Comptonization ($Comptb$) component, 
 [{\it COMPTB} is the XSPEC Contributed model\footnote{http://heasarc.gsfc.nasa.gov/docs/software/lheasoft/xanadu/xspec/models/comptb.html},
see \cite{F08}], a soft $Blackbody$ component 
and the iron line component. The $Comptb$ spectral component has the following parameters:
temperature of the seed photons $T_s$,
the energy index of the Comptonization spectrum $\alpha$ ($=\Gamma-1$), 
the electron temperature $kT_e$,   the  Comptonization  fraction $f$ [$f=A/(1+A)$]
and the normalization of the seed photon spectrum $N_{com}$. 

Some observations show significant excess of emission in
the 6 -- 8 keV region occurs 
which one can  attribute to the iron emission
complex.
The presence of the iron line 
features provides  the evidence for an additional
{\it reprocessed} component in the spectrum. 
It should be noted that 
a subset of observations can be fit without adding the iron line ($\sim$30\%), while the remaining observations require 
the line component including: either two narrow iron $K_{\alpha}$ emission lines ($Gaussians$) or one redskewed iron $K_{\alpha}$ emission 
line (modeled by the $Laor$ XSPEC model). Model fits for several spectra show that the 
 $Laor$-line component is more suitable than the sum  of two $Gaussians$. 
For these reasons,    
an iron K$_{\alpha}$-line ($Laor$) component
(Laor 1991) was included in our model spectrum and applied to all spectra.  
The {\it Laor} model parameters are the line energy, $E_L$, the emissivity index, a dimensionless inner disk radius, $r_{in}=R_{in}/R_g$ (with $R_g=GM_{BH}/c^2$), inclination, $i$, and the normalization 
of the line, $N_L$ (in units of photons cm$^{-2}$ s$^{-1}$).  Note $R_g$ is the gravitational radius of a black hole, with $G$ and $c$ as the common physical constants and $M_{BH}$ as a BH mass.
For the {\it Laor} component we fixed the outer disk radius to the default value of 400 $R_g$ 
while  we vary all the parameters allowed to be free. We also fixed the emissivity index to 3. The inclination is  constrained to a value $i\sim 70^{\circ}$, which is  estimated using scaling technique  (see \S 7.3). 

In turn, to fit the data in the 1 -- 4 keV  range, we use a {\it blackbody} component for  which 
parameters are the normalization $N_{BB}$ and color  temperature $kT_{BB}$. 
In the model we also include  interstellar absorption with a column density $N_H$.

During the fitting procedure we fix certain parameters of the $Comptb$ component. First, we put 
$\gamma=3$. Note the low energy index of the seed blackbody spectrum  is  $\gamma-1=2$. 
We also fixed the value of the   $Comptb$ parameter $\log(A)$  to 2 when the best-fit values of $\log(A)\gg1$  
a  Comptonization fraction $f=A/(1+A)$ equals 
approximately to 1 and the model fit becomes insensitive to the parameter.
We use a value of the hydrogen column density  $N_H=7.7\times 10^{22}$ cm$^{-2}$, which was found by~\cite{Dieters00}.  
As a result  we obtain a satisfactory 
agreement  with  this 
model for both {\it RXTE} and {\it Beppo}SAX  
for all available observations. 



We show examples of fits of  X-ray spectra   using our spectral model in  
Figs.~\ref{BeppoSAX_spectra}--\ref{sp_compar_SAX}  (for $Beppo$SAX data)
and in Figs. \ref{sp_rxte_evol}-- \ref{sp_rxte_compar}  (for   {\it RXTE}  data). 
Spectral analysis of {\it Beppo}SAX and  {\it RXTE}  observations indicates  that X-ray  spectra of 4U~1630--47 can be  satisfactory  
fit 
by the model where its Comptonization component is present by the {\it Comptb} model. 
Moreover,  for the broad-band {\it Beppo}SAX  observations this spectral model  component 
allows to describe  the photoelectric absorption at low energies ($E<3$~keV) 
and the structure of  Fe K band over  6 -- 7 keV energy band in detail. 

\subsection{{\it Beppo}SAX data analysis}


In Figure~\ref{BeppoSAX_spectra} we demonstrate three representative $EF_E$ spectral  diagrams  (green lines) for different states of 4U~1630--47. 
Data are taken from $Beppo$SAX observations 
20114002 ($left$ panel, ``S2'' data set, HSS),
20114005 ($central$ panel, ``S5'' data set, IS),
and 70821005 ($right$ panel, ``S7'' data set, LHS).
The data are shown by black crosses and  
 the spectral model components are displayed  by dashed red, blue and  purple lines for the $Comptb$, 
 $Blackbody$ and $Laor$ respectively. Yellow shaded areas demonstrate an evolution of $Comptb$ component 
during the state transition 
between the HSS ($S2$) and LHS ($S7$)  when the normalization parameter $N_{com}$ of the Comptonization component
 monotonically decreases from 
13 to 2$\times L_{37}/d_{10}^2$
 erg/s/kpc$^2$.
In the {\it bottom panels} we show  the corresponding $\Delta \chi$ vs photon energy (in keV). 

The best-fit model parameters for the HSS state ($left$ panel, $S2$) are 
$\Gamma$=2.62$\pm$0.05,  
$kT_e>$230 keV and $E_{line}$=6.42$\pm$0.04 keV [$\chi^2_{red}$=1.16 for 127 d.o.f], 
while the best-fit model parameters for IS state ($central$ panel, $S5$) are 
$\Gamma$=2.03$\pm$0.03, 
$kT_e$=58$\pm$1 keV and $E_{line}$=7.08$\pm$0.06 keV [$\chi^2_{red}$=1.18 for 127 d.o.f]; and, finally,  
the best-fit model parameters for LHS state ($right$ panel, $S7$) are 
$\Gamma$=2.01$\pm$0.02, 
$kT_e$=160$\pm$50 keV and $E_{line}$=5.38$\pm$0.05 keV [$\chi^2_{ref}$=0.92 for 127 d.o.f]. 

{The {\it blackbody} temperature $kT_{BB}$, weakly depends  on the source state and is consistent with 0.7 keV 
(at a 2-$\sigma$ level of confidence). 
Adding this low temperature $blackbody$ component significantly improves the fit quality for 
the {\it Beppo}SAX spectra. For example, the best-fit 
for the LHS events (id=70821005) is characterized by  reduced $\chi^2_{red}$ of 3.7 (129 d.o.f.) for the model 
{\it without this  low temperature blackbody component}, while  $\chi^2_{red}$  is 0.92 (127 d.o.f.) for the model 
{\it with the low temperature blackbody component}. 
We also find that the seed temperatures $kT_s$ of the $Comptb$ component  varies only slightly  around 1.6 keV. 
A systematic  uncertainty  of 1\% has been  applied to all  analyzed {\it Beppo}SAX  spectra (see more details in Table 3).

The general picture of the LHS-IS-HSS transition is illustrated
in Figure~\ref{sp_compar_SAX} where we put together spectra of the LHS, IS and HSS,
to demonstrate the source spectral evolution from the high-soft to low-hard states based on the $Beppo$SAX observations. 
We should  point out the fact that 
the HSS and IS spectra are characterized by a strong soft {\it blackbody} component and a power law extending up to 200 keV, while 
in  the LHS spectrum, the Comptonization component is dominant and the {\it blackbody} component is barely seen. 

For the {\it Beppo}SAX data 
(see Tables 1, 3) we find that the spectral index $\alpha$ monotonically 
increases from 1 to 1.7 (or the photon index  $\Gamma$ from 2 to 2.7),  
when the normalization  of $Comptb$ component (or mass accretion rate)  changes  by factor of 8. 


\subsection{{\it RXTE} data analysis}

For {the \it RXTE} data analysis we use the information which we obtain
using  the $Beppo$SAX  best-fit spectra.
Specifically, because of {\it RXTE}/PCA detectors cover energies  
above 3 keV, for our analysis of {\it RXTE} spectra we fix a key parameter of the {\it blackbody} component 
($kT_{BB}=$0.7 keV) obtained as a mean value  of 
$kT_{BB}$ 
for the {\it Beppo}SAX spectra.
In Figure~\ref{sp_rxte_evol} we show representative spectra  of 4U~1630--47 for  the LHS, IS, HSS, and VSS. 
Data are taken from {\it RXTE} observations 30172-01-18-12 ($\Gamma=1.6$, LHS), 30172-01-04-00 ($\Gamma=2.2$, IS), 80117-01-05-00 ($\Gamma=3.0$, VHS), 
and 10411-01-03-00 ($\Gamma=2.0$, HSS). Here data
are denoted by black points with error bars.  The spectral model is presented by {\it blackbody}, 
{\it Comptb}, and {\it Laor} components  shown by blue, red, and purple lines 
respectively.

The best-fit model parameters for the LHS state ($top~left$ panel) are 
$\Gamma$=1.62$\pm$0.03,  
$kT_e$=49$\pm$6 keV and $E_{line}$=5.86$\pm$0.06 keV [$\chi^2_{red}$=1.00 for 93 d.o.f];  
while the best-fit model parameters for the IS state ($bottom~left$ panel) are 
$\Gamma$=2.15$\pm$0.02, 
$kT_e$=41$\pm$5 keV and $E_{line}$=6.40$\pm$0.02 keV [$\chi^2_{red}$=1.35 for 93 d.o.f]; 
in turn, the best-fit model parameters for the VHS state ($top~right$ panel) are 
$\Gamma$=2.98$\pm$0.01, 
{ 
$kT_e>$ 200 keV 
}
and $E_{line}$=6.44$\pm$0.07 keV [$\chi^2_{red}$=0.97 for 93 d.o.f]; 
and, finally,  
the best-fit model parameters for the HSS state ({\it bottom-right} panel) are 
$\Gamma$=1.9$\pm$0.8, 
$kT_e$=59$\pm$1 keV and $E_{line}$=6.41$\pm$0.05 keV [$\chi^2_{ref}$=0.65 for 93 d.o.f]. 

Thus {\it RXTE}  
observations cover 4U~1630--47 in four spectral states and reveal 
its spectral evolution from the low-hard 
to high-soft 
states.  
In Figure~\ref{sp_rxte_compar} we illustrate the spectral evolution  for
LHS-IS-VSS-VHS transition using  six representative $EF_E$ 
spectral diagrams which are related to these spectral states in 4U~1630--47.
Similarly to our {\it Beppo}Sax spectral analysis, we use as the best-fit model 
$wabs*(Blackbody+Comptb+Laor)$
for the {\it RXTE} spectral modeling. 
The  data are taken from {\it RXTE} observations 30172-01-18-12 ($green$, LHS), 10411-01-18-00 ($orange$, LHS), 
30172-01-07-00 ($pink$, IS), 70417-01-03-00 ($red$, IS), 10411-01-03-00 ($blue$, HSS), 80117-01-03-00G ($black$, VHS). For clarity of the presentation we use
 the normalization factors of 0.5 and 0.1
for 10411-01-18-00 and 30172-01-18-12 spectra  ($orange$  and $pink$ curves respectively). 
{
An example of the
} 
best-fit parameters of the source spectrum and values of $\chi^2_{red}$ including the number of degrees
 of freedom  for {\it RXTE} spectra is presented in Table 4 
{
for 1998 ($R2$ set).
} 
In particular,  we find  that the illumination fraction $f$ of the  $Comptb$ component varies in a wide range between 0 and 1,
 and  that it 
 undergoes sudden changes during an outburst phase 
for all observations. 
  
In general, for the VHS events (see $blue$ line in Fig. \ref {sp_rxte_compar})
 are 
characterized by a dominant soft {\it blackbody} component and relatively weak power law component with respect to that in the IS and VSS.  In the IS and VSS spectra the contribution of 
the {\it blackbody} component is less than that for the VHS.
In the LHS spectrum, 
the Comptonization component is  dominant and the {\it blackbody} component is weak in agreement with our the BeppoSax analysis. 
An evolution 
between the low state and high state is accompanied by 
a monotonic increase of the normalization parameter
$N_{com}$ of the Compton component  from 0.1 to 30$\times L_{37}/d_{10}^2$ 
erg/s/kpc$^2$ 
and  by an increase of   the photon index $\Gamma$ 
from 1.5 to 3, 
(see  Figure~\ref{saturation}). 
We should point out a clear anti-correlation between  the illumination fraction and the photon  index,
identified   by Stiele et al.  (2013) (combine their Figures  4, middle panels) for  XTE J1650-500.  
Our results show  that for 4U 1630-47  the illumination fraction $f$ and the index  $\Gamma$  also anti-correlate 
in a similar manner [compare 
Fig. ~\ref{saturation} and Fig.  4 in Stiele et al.  (2013)].
 
 
T05 reported a strong bump at $\sim$20 keV in some {\it RXTE} 
spectra of 4U~1630--47, which cannot be satisfactory fit with their adopted model
 [{\it wabs*(diskbb+cutoffpl\-+Gaussian)*smedge}]. 
In our  investigation we  find small 
positive excess at 20 keV for only one particular {\it RXTE} observation (id=50135-02-03-00) studied by  T05.


Previous analyses of other X-ray binaries occasionally show, 
a clear bump at 20 keV in their IS spectra.   
We suggest that it is difficult to explain this feature purely in terms of Compton reflection
because the photon index $\Gamma>2$ and thus  there is a lack of  
photons in the incident spectrum to effectively form this 20 keV reflection bump. In fact, 
this feature can be more naturally attributed to the 
reprocessing of high energy photons into lower energies 
due to  down-scattering off relatively cold electrons in the surrounding disk or 
cold outflow if $\Gamma$ is less than 2 (see Basko et al. 1974 and Laurent \& Titarchuk 2007). 
Laurent \& Titarchuk (2007) demonstrate using a Monte Carlo simulation 
and theoretical arguments that the reflection bump never appears in the emergent spectra if the photon index of the spectrum $\Gamma$ is higher than 2.  The resulting spectrum becomes steeper 
and spectrum is deformed. 

An additional  thermal  component with a relatively 
high color temperature of about 4 -- 5 keV (so called a {\it high-temperature blackbody} component, see e.g., TS09; Seifina \& Titarchuk 2010, ST10 hereafter; Koljonen et al., 2013; Mineo et al., 2012)  is 
 a  component of the spectral  model   
which fits the data.
TS09 
find that eight  IS and  VHS spectra of GRS~1915+105  that require this {\it high-temperature blackbody} component.  These cases are rather related to epochs of radio loud events. Recently, Mineo et al. (2012) find that spectra
in the ``heartbeat'' state of GRS~1915+105 can be fit with models including a {\it high-temperature  blackbody} component of the color temperature 3 -- 6 keV. Furthermore, ST10 
reveal 
24 IS spectra from SS~433 during
radio outburst decay events that had  also  a strong {\it high-temperature  blackbody} component
with color temperature of 4 -- 5 keV. 
One can suggest  that this high-temperature thermal component 
is indicative of the neutron star. However it is well-known that a NS color temperature is in the range $1-2.5$ keV [see e.g.  \cite{lpt97}].
TS09 argue that this {\it high-temperature blackbody} bump could
be a result of the gravitationally redshifted annihilation line
which is initially formed in the close vicinity of a BH horizon.

We should note that a low excess at 10 -- 20 keV in the spectrum residuals for 4U~1630--47 can be excluded 
(in terms of $\chi^2$ criterion) 
if  we  increase $kT_s$ and $kT_{BB}$ temperatures from 0.6 keV and 1.2 keV  correspondingly to 0.7 keV and 1.4 keV changing  the normalization parameters.
Thus  we do not find any significant difference in $\chi^2$ with or without  
a {\it high-temperature blackbody component}.
Both models  show acceptable values of $\chi^2_{red}\sim1$. 
As a result we decide to  apply 
an additive model  $wabs*(Bbody + Comptb + Laor)$ which uses fewer  spectral components.





\subsection{Discussion of the data analysis and  X-ray spectral modeling} 
                    
Our spectral model applied to the data from {\it Beppo}SAX and {\it RXTE} shows  robust  performance throughout all data sets.
Namely, a value of the reduced
$\chi^2_{red}$
is  less or around 1.0 for most observations. For a small 
fraction (less than 3\%) of the analyzed spectra with high counting statistics
$\chi^2_{red}$ reaches 1.5. 
Note that the energy range for the cases, in which we obtain  the  poor fit statistic (two among 425 spectra with $\chi^2$=1.55 for 93 d.o.f),   are related to the iron line region.  

We remind  the  reader that the iron emission feature at 6 -- 8 keV is clearly detected in $low$ (luminosity) states and barely seen in 
$high$ (luminosity) states of 4U~1630--47. Some previous studies [\cite{Ooster98}, \cite{Trud01}] 
did not utilize any iron emission line component. Furthermore, 
{
Tomsick et al. (1998) pointed out absorption lines 
from highly ionized H-like and He-like iron lines, which possibly indicate on the 
presence of  a highly ionized  disc or a wind (see also 
Kubota et al, 2007). 
Generally, for sources at a higher inclination,
the obscuration of 
the continuum from the  central source by the disk rim allows the detection of wind or the photo-ionized  atmosphere due to observations of  narrow emission lines, which are not visible at lower inclinations  due to the stronger continuum emission. Recently, Trigo at el. (2013)  detected highly-ionized  absorption lines from Fe~XXV and Fe~XXVI in XMM-$Newton$ observation of 4U~1630-47. They   discussed a possible origin of these lines in terms of accretion disc wind. In turn,  Ponti et al. (2012) suggested that such winds can significantly affect on resulting X-ray emission of BHs  
particularly for high-inclined BHC binary systems. In addition, 
}
\cite{Cui00} detected a double structure of emission feature fit by two $Gaussian$ lines centered around 
5.7 and 7.7 keV  using the additive continuum model consisting of  {\it multicolor disk} 
(``diskbb'') and {\it powerlaw} utilizing  {\it RXTE} data. However, we do not find any evidence of double line structure in X-ray  spectrum of 4U~1630--47 in the framework of our model [$wabs*(Bbody + Comptb + Laor)$]. 
 
 

It is important to point out  that we find similar 
best-fit model parameters with  
those presented in the literature   for the same 
$Beppo$SAX observations. 
  In particular, the photon index  $\Gamma$, estimated by \cite{Ooster98} 
   for observations $S1-S4$, 
is about 2 -- 2.7 
(both for two additive models: a $powerlaw + disk~blackbody$ model and for a $pexriv + disk~blackbody$ model).
This  similarity 
of the index values to our results  using  these different models is indicative of a correct 
approach for  X-ray spectrum modeling and 
spectral state evolution scenario. 

Thus, our $Bbody+Comptb+Laor$ model shows a good performance when we apply it to  the   the observed spectra from 4U 1630-47.
In particular, LT99 argue 
that Generic Comptonization spectra (GCS) can be formed 
as  a result of the combined $thermal$ and $bulk$ Comptonization effects. 
The difference between $Comptb$ and GCS is in the way they explain 
of the exponential cutoff of the spectrum, which 
is determined by electron energy $kT_e$ in the $Comptb$ case  and by the plasma energy 
(thermal plus bulk) 
in the GCS case. Thus
the $Comptb$ model is applicable to
GCS spectra profile with the warning 
that the meaning of $kT_e$ as the electron temperature can be generalized 
 to include possible effects of bulk-inflow (dynamical) Comptonization.
 In other words, the cutoff energy  can be dictated by the combined effect of thermal and dynamical Comptonization and the best-value value of $kT_e$, in the framework of {\it Comptb} component, indeed gives  the mean value of the thermal and bulk energy.  
 On the other hand the  energy spectral index 
$\alpha$~(or the photon index $\Gamma=\alpha+1$)  is a measure of Comptonization efficiency.
Indeed, TS09 show that the index $\alpha$ is an inverse of the Comptonization parameter $Y$ which is a product of average energy change per scattering, $\eta=<\Delta E>/E$ and the number of scattering $N_{sc}$ in the medium (Compton region).    

We identify, using the
{\it Beppo}SAX observations of  4U~1630--47,
the model of the spectrum which covers  the broad energy band from 0.3 to 200 keV.
On the other hand, the  {\it RXTE} extensive observations 
give  us  an opportunity to  
investigate the overall evolution pattern of the source behavior during all spectral transitions 
in the 3 -- 200 keV energy range.





\section{Overall pattern of X-ray properties \label{overall evolution}}

Observations of 4U~1630--47 show a variety of 
different states, which  are associated with the changes of X-ray energy 
and power spectra. 
Previous investigations of 4U~1630--47 (e.g., Tomsik et al., 2005, Dieters et al., 1999) highlighted 
various outburst behavior and also pointed out   differences in spectral and timing properties 
as well as light curve patterns during individual outbursts. 
However,  Tomsik et al. and  Dieters et al.  analyzed 
individual outbursts applying different spectral models, which can introduce 
corresponding differences in the inferred spectral properties of 4U~1630--47. 
Therefore, because we are interested in the general character of the spectral evolution of 4U~1630--47, we  perform a uniform analysis
applying  the same spectral model to all observations of 4U~1630--47. 
As a first step, 
we construct  a hardness-intensity diagram of 4U~1630--47, 
which  helps us to understand how 4U~1630--47 evolves between different spectral states. 
 Note, that {\it  the hardness (observable flux ratio) is a  measurable quantity, 
 and   directly related  to the index of the spectrum} (see below).




To study the properties of 4U~1630--47 during the spectral transitions, 
associated with  significant changes in source luminosity, 
we  investigate  the direct observational dependence  
of hard color [10-50 keV/3-50 keV] (HC) on   the 3-10 keV  flux 
measured in units  of $10^{-9}$ erg s$^{-1}$cm$^{-2}$ [hardness-intensity diagram (HID)].
In Figure~\ref{HID_6obj} we present the flux ratio HC 
versus  the 3-10 keV  flux measured in units  of $10^{-9}$ erg s$^{-1}$cm$^{-2}$ 
based on  the {\it RXTE} data ({\it black triangle} points for 4U~1630--47 there). 
As we see from this Figure the hardness of the spectra monotonically decreases when this energy flux 
in the 3-10 keV band increases. In other words the spectrum becomes essentially softer for higher fluxes. 
The hardness ratio saturates at high flux values, an observational fact is indicative  of the spectral evolution of the source 
from the LHS through the IS towards  the HSS.


As the source moves from  the LHS towards  the IS  the hard color (HC)  drastically drops from 3.7 to  0.1 while the 3-10 keV flux only slightly increases.  When the source further moves  from   the IS towards  the soft states (VHS and HSS) the HC saturates while  the 3-10 keV flux increases factor  by 10, i.e. from 2 to 20$\times 10^{-9}$ erg s$^{-1}$cm$^{-2}$. 
In Figure~\ref{HID_6obj}  we  compare  this HC behavior versus the flux for  4U~1630--47 with that observed in other BHCs and NSs

We use the HC [10-50 keV/3-10 keV] versus  flux in the 3 -- 10 keV range  
in the form of 
HID for six sources: 
4U~1630--473 (BHC, $black$), SS~433 (BHC, $crimson$), 
4U~1820-30 ($atoll$ NS, {\it bright blue}), GX~3+1 ($atoll$ NS, $green$) and 4U~1728-34 ($atoll$ NS, $blue$)  
and GX~340+0 ($Z$ type NS, $red$). As one can  see from  this Figure,
HID for  4U~1630--47 exhibits two separate branches related to the $hard$ spectral 
states (vertical branch) and to the $soft$ states  (horizontal branch),  
whereas all NS sources (except 4U~1728-34) show only the  {\it soft}  horizontal branch elongated close to 
the soft branch of 4U~1630--47. 
A unique stability (or quasi-constancy) of 
the hardness ratio for NSs versus  the $3-10$ keV flux is an observational demonstration 
for the  stability of the photon index $\Gamma$ established for a number of  NS sources 
(see Farinelli \& Titarchuk 2011, Seifina \& Titarchuk 2011, 2012 and Seifina et al. 2013).  
This hardness ratio vs the  $3-10$ keV flux provides a unique 
diagnostics for  the nature of a given compact object.  

As evident from this Figure, 4U~1630--47 shows both wider luminosity range and a wider hardness ratio 
range than any NS source. 
This observational features can  be related to different rates of 
mass transfer in  NS and BH systems. 
Thus, the comparison of HIDs, in principle allows one to probe physical properties of compact 
objects, in particular, their possible nature, directly related to a shape and localization of 
HID tracks.


\section{Evolution of X-ray  spectral properties during spectral state transition
\label{evolution}}
A number of X-ray spectral transitions of 4U~1630--47 have been detected 
by {\it RXTE} during 1996 -- 2004 ($R1$ -- $R7$  sets).
We have searched for  common
 spectral and timing features which can be revealed during these spectral transition events.

 The X-ray light curve of 4U~1630--47 shows  complex 
 behavior in  a wide range  of time scales: 
from seconds to years  (e.g., Belloni et al., 1999; T05).
Here  we discuss  the source variability on the time scales of   hours. 
In Figure~\ref{evolution_all}  we demonstrate 
 the source  and model parameter evolution for 
all analyzed 
outburst spectral transitions. 
As one can see from  the flux 
panel (second  from the top),
all outbursts of 4U~1630--47 are characterized 
by a  significant increase of the 3$-$10 keV  flux. Some events  also demonstrate an 
increase of flux in the 10$-$50 keV energy band (see particular events at MJD 50860 and 52260-52290). 
For such outbursts   with a good rise-decay coverage, 
the predominance of the 10 -- 50 keV flux over the 3 -- 10 keV flux can 
be related 
to a moderate mass accretion rate regime. In fact, 
the spectral index $\alpha$ 
is  less   in these dates (at MJD 50860 and 52260-52290)
than that for 2002-2004 (MJD  52520-53080) events
(see  Fig. 
 \ref{evolution_all}).  
The {\it pivoting effect} in 3-10 keV/10-50 keV flux relation  corresponds to the spectral state change. 
Namely, in the framework  of our spectral model (see \S \ref{spectral analysis}) these spectral transitions  
are clearly seen as a  decrease of the normalization
$N_{Com}$ 
and  a spectral index  decrease  (see Fig.~\ref{evolution_all},  two bottom panels at MJD 50860 and 52260).

The spectral index $\alpha~(=\Gamma-1$)  is well traced by soft X-ray flux (compare  the $bottom$ and two upper panels of  Fig.~\ref{evolution_all}).
Moreover,  we do not find  any  difference in  the correlations  
between $\Gamma$ and  soft X-ray flux  for different outbursts
 which was previously  pointed out 
by a number of  authors [e.g., T05, Trudolyubov et al., (1998)]. 
We can also suggest that  the 4U~1630--47 spectral state evolution can be traced 
by the illumination fraction $f$. 
For example, an excess of  the 10$-$50 keV flux over that for  the 3 -- 10 keV energy band 
(at the low/hard and intermediate states)  mostly
 occurs   for high values of $f$  ($0.5<f<1$). 

The actual difference between outbursts is related to  
the peak flux reached during an outburst which 
probably depends   
on 
the value of $\dot M$.  
The outbursts  can be mainly  distinguished  by saturation levels  achieved at the maximum of individual outbursts (see low panels of Fig.~\ref{gam_t_e}).
Higher saturation levels of the index $\Gamma$ correspond  to higher mass accretion rates. However the observed index saturation  level of $\Gamma$ never exceeds 3 as  predicted by  the theoretical calculations of  Titarchuk \& Zannias (1998). 
All  superimposed index-normalization
correlations  can be seen in
a joint diagram in Fig.~\ref{1655_scal} which 
provides  a generic picture of  spectral (index) evolution  vs $\dot M$ through 
all outburst activity of 4U~1630--47. We should notice that  all outbursts of  4U~1630--47 are associated with {\it low} radio    
activity except of the 1998 outburst which  is  accompanied by a strong radio emission [see a possible explanation of the radio activity  by \cite{fend06} and \cite{corb13}].
We can suggest that during the 1998 outburst  the total mass accretion rate from the companion was so high that some  part of this mass flux went to the outflow leading to a strong radio emission  [due to high radiation pressure, see  an explanation of this effect in \cite{tsa07}] and another one  accretes through the disk into the black hole.

Using $Beppo$SAX data and  our model, $wabs*(Bbody+Comptb+Laor)$  
 we  identify a  
{\it blackbody} component  which 
has the temperature $kT_{BB}\sim$0.6 -- 0.7 keV and the seed blackbody temperature $kT_s\sim$1.2 keV. 
{\it RXTE} data show that $kT_s$  varies in the 1.0 $-$2.0 keV interval 
while  $kT_{BB}$ is fixed at 0.7 keV.  These best-fit values of $kT_s$ are different from  those obtained by T05 using the same {\it RXTE} data (herein the $R5$ set). 
Namely, T05 find that $kT_s$ changes  in the 2.7 -- 3.8 keV interval for about half of VHS spectra of 4U~1630--47, while the other half have these  temperatures less than 1.8 keV 
for the model $wabs*(diskbb+powerlaw)$. 
Using  the same spectral model 
\cite{Trud01} obtain $kT_s$ in the range of 1.2 -- 1.8 keV for $R2$ set. 
This difference  of the seed photon  temperature values found by T05  and Trudolyubov et al.  can be explained by  particular properties of the spectra  for $R2$ and $R5$ outbursts. However, we show, using our spectral model, that  $kT_s$ values are similar for all outbursts 
and change between 1 and 2 keV.
We can suggest that this  difference  between T05  and our 
results  is possibly  due  the  different spectral models used.


\section{Correlations between spectral and timing properties during spectral 
 state transitions \label{transitions}}

The {\it RXTE}~ light curves have been analyzed using the FTOOLS {\it powspec} task. 
The timing analysis {\it RXTE}/PCA data was performed in the whole PCA energy
 range by combining the high time resolution {\it event} and {\it binned} data modes.  
We have 
generated power density spectra (PDS) in  0.1 -- 512 Hz frequency range
with milisecond time resolution. We subtracted the noise contribution due to
Poissonian statistics. 
To investigate 
the evolution of the source timing properties, we modeled 
the PDSs  using  
analytic models and the $\chi^2$ minimization technique in the framework 
of QDP/PLT plotting package\footnote{http://heasarc.gsfc.nasa.gov/ftools/others/qdp/qdp.html}.



The broad-band power density spectrum 
of the source is commonly
presented by the noise component, which shape in the LHS and IS usually has
a band-limited noise (BLN) profile 
approximated by an empirical model $P_X\sim [1.0+(x/x_{*})^2]^{-in}$
[the $KING$ model in QDP/PLT].
The parameter $x_*$ is related to the break frequency and 
$2\cdot in$ is a slope of the PDS continuum after the break frequency  $x_*$. 
In addition to the flat-top continuum in the LHS and the IS 
PDS often shows quasi-periodic oscillations (QPOs), modeled 
by Lorentzians. 
In the HSS and VHS  an additional power-law component (very low
frequency noise, VLFN) is needed to fit the data at the lowest
frequencies.


In Figure~\ref{pds_sp_1998} we show  details of the evolution of
X-ray timing and spectral characteristics for the 1998 {\it rise} transition ($R2$). 
At the $top$ of this Figure  
we show an evolution of the 1.3 -- 12 keV ASM flux 
during this outburst. We   choose representative observations designated
by $red$/$blue$  colors and letters A, B and C to 
demonstrate  the PDS for six characteristic  epochs    
at MJD =  50853.1/50855.8, 50856.1/50857.8 and 50862.6/50864.2, respectively,
as shown on the ASM flux diagram at the top of Fig.~\ref{pds_sp_1998}.


On the $left$ $bottom$ panel of Fig.~\ref{pds_sp_1998} we plot PDSs for three observations indicated as A, B, C moments, while  the corresponding energy 
spectra in the form of $E*F(E)$ diagram are shown on the $right$  $bottom$ panel.
In the  energy spectral diagrams the data points are shown by black  while
 the spectral model components are presented  by {\it red}, {\it blue} and {\it pink} 
dashed lines for $Comptb$, $Blackbody$ and $Laor$  components respectively.

Points A [{\it red} (30178-01-01-00) and {\it blue} (30178-01-03-00)] correspond  to the  IS, whereas points B and C
[B {\it red} (30188-02-02-00), B {\it blue} (30178-02-02-00), C {\it red} (30178-01-10-00), C {\it blue} (30188-02-14-00)] are related to the VHS.
All of these observations, except  30188-02-14-00 (C $blue$), exhibit power spectra with strong BLN  accompanied by  QPO peaks.
Evolution of the temporal properties of
the source during the {\it rise} phase of the 1998 outburst is characterized by a
monotonic decrease of the total rms amplitude (from 28\% to 10\%) 
and an increase of the BLN break and QPO centroid frequencies (from 2 Hz to 13 Hz).  
 Note that during X-ray flux plateau intervals the source emission 
is also characterized by the stable spectral and timing parameters.
The 1.3$-$12 keV flux rise is accompanied by  a systematic shift of the frequency of maximum power in PDS 
  towards higher frequencies (see corresponding PDS$\times$frequency diagrams  for A and B points). 

The energy spectra shown in the {\it right} column 
(panels A2, B2, C2) are related to the corresponding power spectrum diagrams,  for panels: 
A1 (point A {\it blue}), B1 (point B {\it blue}), C1 (point C {\it red}). $E*F(E)$ diagrams  
demonstrate an evolution of the spectral properties of 4U~1630--47 during the outburst rise phase, 
which  is characterized by  a monotonic steepening  
of the photon index $\Gamma$ from 1.5 to 3, (see {\it red} points in Fig.~\ref{gam_qpo_freq_scal}). 
Note that similar correlations were revealed in many X-ray BH binaries (e.g., ST09, TS09, TSA07 and SBKM13). 
This simultaneous spectral and timing analysis shows that state transitions in 4U~1630--47 are in agreement
 with the {\it canonical} 
BH spectral state evolution (see e.g. ST09).
Namely, the correlation of the photon index  $\Gamma$ with  the QPO frequencies  (see e.g. ST09, SBKM13 and presented work) ) can be considered as  observational  evidence for the presence of the transition layer (TL).
In the framework  of this scenario these QPOs are caused by  oscillations in this  bounded TL with
frequencies equal to eigenfrequencies of the configuration.
Any boundary configuration, in our case  the TL or Compton cloud,   is characterized by its own eigen-frequency. In order to observe this frequency there should be enough power in excitation of this frequency. It is not by chance we see these low frequency QPOs situated  very close to the break frequency in the power spectrum [for more details on  the theory of oscillations of a bounded medium, see Landau \& Lifshitz (1976) and also  Titarchuk, Shaposhnikov \& Arefiev (2007) (TSA07)]

Consequently, 
this evolution 
of the QPO  frequency  can be 
interpreted as 
a change of  the TL size, or the inner  boundary of the   Keplerian accretion disc where the disk starts to adjust to sub-Keplerian motion in the TL.  Thus, the X-ray spectrum  is formed in the relatively hot TL    and evolves as a result of thermal and dynamical Compotonization of soft (disk) photons  and QPOs are seen as intrinsic oscillations  of  the TL zone.  

While the exact nature of QPOs is under debate the prediction of the TL oscillation interpretation
that the QPO frequency increase  with the flux
is confirmed by the observations. Moreover, the spectral index vs QPO correlation established for many BHC sources  allows us to estimate the compact object masses in these sources which we find to be  consistent with mass function estimates (see ST09).  
We argue that 
that  
the observable correlations between  spectral and timing characteristics seen in  4U~1630--47
are consistent with that obtained in other BHCs (see e.g. ST09 and SBKM13).


\section{Discussion \label{disc}}

\subsection{Saturation of the  index is a possible signature of a BH  \label{constancy}}

In our analysis of the evolution of the photon index $\Gamma$  in 4U~1630--47 
using  {\it RXTE} and {\it Beppo}SAX observations we have firmly established that the index 
saturates with  the low-QPO frequency and probably with the $Comptb$-normalization $N_{com}$  which is proportional to the (disk) mass accretion rate, $\dot M$ (see Fig. \ref{1655_scal}).  
ST09 give strong arguments that this index saturation is 
a signature  of converging flow onto a BH. 
In fact, the Comptonization spectral index is an  inverse of a product of the number of scatterings $N_{sc}$ and the efficiency of upscattering $\eta$.  In the converging flow $N_{sc}$ is proportional to the mass accretion rate $\dot M$  and $\eta$ is inversely proportional to $\dot M$ for $\dot m =\dot M/\dot M_{\rm Edd}\gg1$ and thus one can expect the index saturation when $\dot m\gg 1$.
In addition, we reveal  that the index $\Gamma$ correlates and saturates when QPO 
frequency increases.  The  index-QPO relation now established in many BHs, strongly suggests  that low-frequency QPOs are   a result of an oscillatory process in the Compton cloud (TL).


  \cite{tlm98} predicted that  the TL should become more compact when mass accretion rate, $\dot M$ increases.  
For a BH case \cite{LT99}, (2011), hereafter LT99 and LT11,
find, using Monte Carlo simulations,  that the index  should saturate 
when $\dot M$ increases. 
The photon index increase and subsequent saturation   versus $\dot M$
was firmly established by ST09, TS09  and ST10 and we confirm  this effect in the current  work.   
In particular,  as one can see in  Figure \ref{1655_scal}
the  values of $\Gamma$ 
monotonically increase  from 1.2  and   finally saturate at a value about 3.  

Thus, we  argue that the X-ray observations of 4U~1630--47 reveal the index 
saturation vs mass accretion rate which can be a signature of the converging flow 
(or a BH presence) in this source. 
The index$-${\it Comptb} normalization (or $\dot M$) and the index$-$QPO the correlations
   allow us  to estimate a  BH mass  in 4U~1630--47 
  (see  Fig. \ref{1655_scal} and  \S~\ref{bh_mass} below) 
even without  an observation of  the optical counterpart of 4U~1630-47.

As an additional argument in favor of  a BH (or event horizon)
presence is  the observational  fact that  the cutoff energy  of the spectra,  $E_{cutoff}$ 
have  values about 200 keV 
[see the upper panel of  Fig.~\ref{gam_t_e} and also \cite{grove98}].  
It is interesting that 
a similar  
result was obtained by LT99 and LT11 using Monte Carlo simulations.
They argue that 
for high mass accretion rates (exceeding the Eddington limit, $\dot m=\dot M/\dot M_{\rm Edd}>1$) the cutoff energy is of the order of the electron rest mass,  $E_{cutoff} \sim a\cdot m_ec^2$, 
where $a\sim 0.5-0.7$.    
Thus the case of $E_{cutoff}\gax 200$~keV indicates a significant effect of the bulk motion inflow, which, in turn,  points 
to  the presence of a BH event horizon.

We also find different spectral index  saturation levels for different outbursts (see $bottom$ panel of Figure~\ref{gam_t_e}). 
Furthermore, complex flaring events during the longest  activity period, 2003 -- 2004 
exhibit a number of close but clearly separated levels. 
The 2003 outburst  (MJD=52681.8 -- 52689.1) is characterized by a low saturation level of $\Gamma$
($yellow$ stars in Figure~\ref{gam_t_e}). 
The index saturation value $\Gamma_{sat}$  
is  probably related to a value of  of the converging inflow electron temperature $kT_e$. Namely,  
$\Gamma_{sat}$ increases when $kT_e$ decreases (see LT99 and LT11). The same behavior was found by ST09 in XTE 1550-564 for multiple state transitions during 1998 outburt. 



\subsection{On the non-monotonic behavior of the cutoff energy $E_{cut}$ versus index $\Gamma$
\label{cutoff_energy}}

Our spectral analysis   reveals a non-monotonic behavior of the cutoff energy  $E_{cutoff}$ versus index $\Gamma$ (see Fig.~\ref{gam_t_e}).  In the upper panel of this Figure
we demonstrate how    $E_{cutoff}$  changes with $\Gamma$.  
At low  values of the index,  ($\Gamma<2$) the cutoff energy  $E_{cutoff}$ decreases   and reaches 
 its minimum  in the 50 -- 80 keV  range.  Then for  $\Gamma\gax2$  the energy $E_{cutoff}$ starts to increase again. 
Similar behavior of $E_{cutoff}$ vs $\Gamma$ was predicted by LT11  based on Monte-Carlo  
  simulations. 



The $E_{cut}-\Gamma$ track 
during the 2000 -- 2002 outbursts, at the 2003 decay and during  the 1996 -- 1999 \& 2002 -- 2004 outbursts  were  different from each other
(see Fig.~\ref{gam_t_e}). 
Namely, 
the saturation levels for 2000 -- 2001 and 2003 ({\it decay}) spectra were  lower 
($\Gamma\sim 2.3-2.4$ and 
$\Gamma\sim 1.8$, 
respectively) than that  for other  spectra for which $\Gamma\sim 3$. 
It is interesting to note that the 2000 -- 2001 data ($green$ points) lie close to the 
1996 -- 1999 and 2002 -- 2004
outburst data for LHS-IS transitions ($red$ points). This indicates that during rise episodes 
similar accretion regimes are operating 
for all outbursts.
Then, 
$E_{cutoff}$ starts to increase and  the index saturates at  different values  of $\Gamma_{sat}$
for each of the  outbursts: $\Gamma\sim 2.3-2.4$ ($green$ stars  for 2000 -- 2001 outbursts), 
$\Gamma\sim 1.8-1.9$ [$yellow$ stars  for 2003 ({\it decay}) outburst]   
and $\Gamma\sim 2.8-3$ ($red$ stars)  for the remaining 1996 -- 1999 and 2002 -- 2004 outbursts
(see  Fig.~\ref{gam_t_e}).
According to  the LT11 Monte Carlo simulations 
different  index saturation levels  are caused by 
different  values of the electron temperature $kT_e$, where higher $\Gamma_{sat}$ correspond to
the lower values of  $kT_e$. In other words at higher mass accretion rate  the plasma temperature $kT_e$ and consequently an  efficiency of thermal Comptonization substantially decreases.
In this case 
the bulk motion Comptonization dominates  over the thermal one forcing $E_{cutoff}$ to be shifted to higher values.  


The $E_{cut}-\Gamma$ dependence shown by 4U 1630--47 is strikingly similar to those discovered 
in GX 334-9 by \citet{motta2009} and  subsequently reported in XTE J1550-564 by \citet{ts10}. 
Especially remarkable is the resemblance of the $E_{cut}-\Gamma$ track shown in Fig. \ref{gam_t_e}
for 4U 1630--47 to that 
for XTE J1550-564 \citep[see Fig. 4 in][]{ts10}. 
Therefore, the presented correlations between cutoff energy and index for 4U 1630--47 
add yet another aspect of similarity of this source with other BHC sources.
 
\subsection{Determination of  BH mass in 4U~1630--47\label{bh_mass}}

In this Section we apply the scaling technique developed in ST07 and ST09 to make a BH mass 
estimate for 4U~1630-47 based on X-ray data. Note that BH mass evaluation for this object is 
impossible by any traditional 
dynamical methods based on optical data because of large visual  extinction towards the source.

We carry out a scaling analysis of 4U~1630-47 with a number of sources such as XTE~J1550-564, H~1743-322 as well as 
GRO~J1655-40 to give  cross-check results. It is worth  noting  that we can proceed with this scaling technique  if the correlations between these particular  sources are self-similar. In other words, we implement this scaling 
technique when the given correlations have: i) the saturation part at high $\Gamma$-level of corresponding track, 
ii) the same index saturation levels and iii) the same slopes as an index  function of QPO frequency. 

\subsubsection{Scaling of 4U~1630-47 with XTE~J1550-564}

The X-ray nova XTE~J1550-564 is a well-studied BHC XRB with one of the best measured mass and distance among  known stellar mass BH sources (see Table 5). Note that the scaling between 4U~1630-47 and XTE~J1550-564 can be implemented using  1998 rise  outburst data of XTE~J1550-564 observed with {\it RXTE}. This 1998 rise outburst was previously discussed in detail by 
many authors [see, for example, Remillard et al. (2002), hereafter R02, and ST09].
 R02 and ST09 implemented  a deep timing and spectral analysis for {\it RXTE} data collected during the rise outburst (time) interval from 08/09/98 
to 16/10/98 (MJD 51069 - 51105).  Remillard et al. 
found the type-C LFQPOs in the power spectra and investigated the behavior of QPO  frequency as a function of the total rms amplitude. 
Casella et al.  (2005) noted that all these  type A, B and C QPOs, in particular, for XTE~J1550-564 forms the corresponding tracks 
at proper areas in the diagram of QPO frequency versus the inverse of the $rms$ (hereafter $S_{rms}$, see $top$ panel on the $right$ 
column of Fig.~\ref{1655_scal}).

 ST09
 revealed  a change of the photon index  $\Gamma$ from 1.4 to 2.9 in the 
X-ray spectra of XTE~J1550-564.
$\Gamma$ increases when the type C LFQPO increases  forming  a characteristic correlation having 
the saturation part at high value (8 -- 13 Hz) of LFQPO (ST09, see also a top panel on  the left column of Fig.~\ref{1655_scal}) with $\Gamma_{sat}\sim 3$. It is 
interesting that the spectral index also shows a positive correlation with the {\it normalization} parameter (proportional to mass accretion rate)   along with the saturation plateau  at the same level $\Gamma_{sat}\sim 3$ at high values of the 
{\it normalization} parameter.   
This behavior of XTE~J1550-564 during outburst rise is similar to the evolution 
of spectral and timing parameters of 4U~1630-47 in terms of $\Gamma$-Norm and $\Gamma$-LFQPO diagrams.
The saturation levels and the slopes of corresponding correlation tracks as a function of QPO frequency and the spectral 
normalization are the same (see top and $middle$ panels of the $left$ column in Fig.~\ref{1655_scal}). 




In order to proceed with the scaling procedure we  parametrize the correlation pattern 
$\Gamma$-Norm/
$\Gamma$-QPO frequency for a reference transition in terms of the analytical function 

\begin{equation}
F(x)= A - (D\cdot B)\ln\{\exp[(1.0 - (N/N_{tr})^{\beta}))/D] + 1\}.
\label{scaling coefficient}
\end{equation}

By fitting this functional expression 
to the correlation pattern, we find a set of parameters A, B, D, $N_{tr}$, and $\beta$ 
that represent a best-fit form of the function $F(x)$ for a particular correlation curve. 
For $N\gg N_{tr}$, the correlation function F(x) converges to a constant value A. Consequently, A is the value of the 
index saturation level, $\beta$ is the power-law index of the inclined part of the curve and $N_{tr}$ is the value 
at which the index transitions, i.e. levels offset. Parameter D determines how smoothly the fitted function saturates 
to A. We scale the data to a template by applying a transform $N\to s_N\times N$ until the best fit is found. 

XTE~J1550-564, 
has a reliable 
dynamical mass 
determination of 9.5$\pm$1.1~M$_{\odot}$, (Orosz et al., 2002), 
which corroborates our scaling method  which gives the value of 10.7$\pm$1.5~M$_{\odot}$ (see ST09). 
We use the $\Gamma-$Norm correlation and the scaling BH mass as a reference data in our subsequent analysis.
The reference data set is well represented by the function defined in Eq. 
(\ref{scaling coefficient}).   We  fit  $F(x)$ 
to best represent the XTE~J1550-564 $\Gamma-$Norm track varying the parameters A, $N_{tr}$, $\beta$. To test for self 
consistency we have used the chi-squared statistical minimization method to determine the best fit parameter values. 
Parameters D and B are not well constrained by the data. Therefore, we fix those parameters at 1.0 and 0.6 
respectively based on our previous experience with correlation parameterization (Table 4 of ST09). The best-fit approximations by the  analytical function (Eq. \ref{scaling coefficient}) are presented 
in Fig.~\ref{1655_scal} along with observational data. 
As a result, for free parameters we obtain the following best-fit values:
A=2.94$\pm$0.08 (saturation level), $N_{tr}=0.011\pm 0.004$, $\beta=0.99\pm 0.14$.   We then 
  obtain the $first$ scale factor $s_N=2.2\pm 0.5$.  

The {\it second} 
scaling factor $s_{\nu}$ is followed from the scaling law 

\begin{equation}
s_{\nu}=\frac{\nu_r}{\nu_t}=\frac{M_t}{M_r},
\label{first scaling law}
\end{equation}

\noindent which based on the inverse dependence of the QPO frequencies on BH mass in terms of the {\it Transition layer} model 
(see details in TLM98; TF04; Appendix A of ST09) using 
the scaling  of the photon index vs QPO frequency correlation. Here $M_r$ is BH mass of the reference source (XTE~J1550-564), 
$M_t$ is BH mass of the target source (4U~1630--47). Subscripts $r$ and $t$ of frequency $\nu$ denote   the reference and target sources, respectively.

As it is seen  from Figure~\ref{1655_scal} (top left panel) the  index-QPO frequency diagram for 4U~1630--47 
({\it blue} points) follows   that for XTE~J1550-564 
({\it red} points). Thus, the  index-QPO frequency diagrams for both sources  
are nearly equivalent. 
In this way, one obtains  that the  
scaling factor  $s_{\nu}=1.00\pm 0.01$ and  the mass of target 
source (4U~1630--47) using Eq.~ (\ref{first scaling law}) is  
$M_t=s_{\nu} M_r$. Assuming the mass estimate of 9.5$\pm$1.1 M$ _{\odot}$ 
(Orosz et al. 2002) we estimate a BH mass 
for 4U 1630-47 (see Table \ref{tab:par_scal}).

Note that $\Gamma$-QPO correlation  is independent of the distances and the geometrical factors 
such as an inclination angle (see Eg.~\ref{first scaling law}), while $\Gamma$-Norm correlation is closely related  to these  parameters. 
Thus using the mass estimate 9.5$\pm$1.1 M$_{\odot}$ for  4U~1630--47   we have  an opportunity to evaluate the inclination angle for this source 4U~1630-47 using  the $\Gamma$-Norm scaling.


According to the scaling laws (see Eqs.~8-9 in ST09), the expression for 
geometrical factor $f_G$ in 4U~1630-47 is:

\begin{equation}
f_G=s_N~\frac{M_t d^2_r}{M_r d^2_t}
\label{geom factor}
\end{equation}
where 
geometric factor, by definition (ST09),  $f_G=(\cos\theta)_r/(\cos\theta)_t$, inclination angles $\theta_r$,  
$\theta_t$ and $d_r$, $d_t$ are distances to the reference and target sources. 
Furthermore,  if we know  mass $M_t$ (see Eq.~\ref{first scaling law}) and the $first$ 
scaling index $s_N$ we can find an inclination angle $\theta_t$ for the target source, 4U~1630--47 using  a formula for  
$f_G=(\cos\theta)_r/(\cos\theta)_t$ and values of 
parameters ($\theta_r$ and $d_r$) for XTE~J1550-564 
\begin{equation}
(\cos\theta)_t\sim \frac{(\cos\theta)_r}{s_\nu s_N}\times \left(\frac{d_t}{d_r}\right)^2.
\label{scaling coefficientm}
\end{equation} 

We show the estimated  values of  BH mass  and the inclination angle with the proper error bars  for the  
target  source (4U1630--47)   
 along with parameters for the 
reference  source (XTE~J1550-564) in Table~\ref{tab:par_scal}.




It is interesting that there are many  indications obtained by other methods in a favor of high inclination value for the binary 4U~1630-47 . Recently, Mu$\tilde n$oz-Darias et al. (2013) give  clear indications on a high 
inclination of 4U~1630 based on the {\it RXTE} observations during outburst transitions. They find 
significant  difference in the shape of the tracks that high inclination systems display in the color/luminosity diagrams  from that of low inclination systems. 


Note that the derived  estimate of the inclination angle 
$\theta\lax 70^{\circ}$ in 4U~1630--47 predicts that 
deep orbital eclipses   should be absent. 
In fact, X-ray observations of the source
do not show any strictly recurrent dip events. 
However, this does not exclude a  possibility of partial orbital eclipses. 
Furthermore, the presence of 
a  wind from the accretion disk can result in 
an absorption of the X-ray  emergent radiation due  to a high optical depth in a wind envelope.
This picture  is consistent with an  absorption line detection in  X-ray high-resolution spectra of 4U~1630--47 
with $Suzaku$~[\cite{Kubota07}]
who  find that 
these X-ray spectra 
reveal strong absorption lines from highly ionized (H-like and He-like) iron K$_{\alpha}$ 
at 7.0 keV and 6.7 keV.  
A possible origin of these  lines can be associated with
an outer part of the   
accretion disk. 


A number of 
X-ray binaries have been 
found to  show   
absorption lines originated  from highly ionized elements 
(Boirin et al. 2004). 
These types of sources  ranges from  microquasars,  e.g. GRO~J1655-40 
(e.g., Miller et al. 2006; Yamaoka et al. 2001; 
Ueda et al. 1998) to LMXB,   e.g. GX~13+1 (Sidoli et al. 2002; Ueda et al. 2004). All of them are viewed at high inclination angles. 

In addition, the detection of a dip in the X-ray light curve of 4U~1630--47 during
the 1996 outburst (see Kuulkers et al. 1998) indicates that the
inclination of the system is fairly high, 
which is similar to GRO~J1655-40.
This single 
dip (140 seconds of time duration) during 1996 outburst (Kuulkers, 1998) could  be caused by  inner conditions such as the attenuation (screening) of X-ray region by inner structures  of the accretion disk, a sporadic (irregular) absorption the clumping  in the wind, by high current  inclination due to disk precession and so on. However, after the 1996 outburst  no more similar dips were detected.

 One can conclude that
 the detection of  a single isolated dip  (possibly partial orbital eclipse) in the 1996 outburst  can indicate a  high orbital inclination angle ($\theta\le 75^{\circ}$).
 On the other hand, the inclination angle is not sufficiently high because  
the regular eclipses are not observed.
This imposes 
constraint on the inclination angle  $\theta\lax70^{\circ}$.

\subsubsection{Scaling of 4U~1630-47 with H~1743-322}

For an  additional cross-check 
of the mass evaluation of $\sim 10 M_{\odot}$ for  BH mass 
in 4U~1630 we use another object, H~1743-322. 
Note that the $\Gamma$-QPO and $\Gamma$-normalization correlation curves for the 
2003 rise data of H~1743-322 (taken from ST09) are  self-similar   with correlations that we 
derive  for 4U~1630-47 (see Fig.~\ref{1655_scal}). 
The 2003 outburst in H~1743-322  was detected on March 21 in hard X-rays (15 -- 200 keV) by INTEGRAL (IGRJ17464-3213; 
Revnivtsev et al. 2003). Follow-up observations with {\it RXTE} led to the recognition that this source is a recurrent X-ray 
Nova which was  first observed with the Ariel V All-Sky Monitor by Kaluzienski \& Holt (1977). 
Further, the X-ray spectral 
and temporal properties of X-ray data as well as radio detection of large-scale relativistic X-ray jets 
(Corbel et al. 2005) allow to  classify   H~1743-322 as a BHC (Fender 2006). A transient X-ray system 
H~1743-322 is a source of X-ray, radio, optical and near-infrared emissions, which enable one  to evaluate its parameters 
such as the distance, orbital inclination and BH mass (McClintock et al., 2007). 

As it is  seen from Table 5,  
H~1743-322 is characterized by almost the same BH mass as that for 4U~1630-47 (based on the mass determination by the scaling with XTE~J1550-564, see  the previous section). 
Indeed H~1743-322 and 4U 1630-47  demonstrate a similar index  saturation level
and $\Gamma$-QPO correlations.
(see Fig.~\ref{1655_scal}).
Considering that a BH mass for 
H~1743-322 ($M_{H1743}\sim 10~M_{\odot}$) we can also confirm our  BH mass estimate 
$\sim 10~M_{\odot}$ for 4U~1630-47.

Moreover, these two sources have the same geometrical parameters (distances and inclination angles). 
In this case their  $\Gamma$-Norm tracks should also  resemble 
each other.
As seen from Figure~\ref{1655_scal}, the correlation tracks for both objects are similar, showing 
the same saturation levels. 
However, $\Gamma$-Norm  correlation for H~1743-322 is located to the right from that for 4U~1630-47. 
We apply Eq.~(\ref{first scaling law}) to $\Gamma$-QPO correlations for 
4U~1630-47 and H1743-322 (bottom right panel of Fig.~\ref{1655_scal}) to obtain $s_{\nu}=0.85\pm 0.08$. We thus estimate that  
$M_{1630}=9.80\pm 0.08 ~M_{\odot}$.  
We can also estimate 
$M_{BH}=9.6\pm 0.1~ M_{\odot}$ in 4U~1630-47  using $\Gamma$-Norm correlations (bottom left panel of Fig.~\ref{1655_scal}) 
if we use $M_{r}$ ($M_{1743}=11.5~M_{\odot}$, McClintock et al. 2007) for the reference source in Eq. (\ref{geom factor})
assuming  
$f_G=1$ (namely, inclinations for both objects are the same,  $\sim 70$ degrees), 
$s_N=1.2\pm 0.1$, $D_{1630}=10$ kpc, $D_{1743}=10$ kpc .






Note 
 that spectral and temporal evolution of H~1743-322 is  similar to that 
observed in  XTE J1550-564 during  1998-1999 (Sobczak et al. 2000b; RM06) and in  GRO J1655-40 during 1996-1997 (Sobczak et al. 1999; Remillard et al. 2002b; RM06). 

\subsubsection{Scaling of 4U~1630-47 with GRO~J1655-40}

The photon index $\Gamma$  for 4U~1630-47 during 2000 -- 2001 period  demonstrates a  lower 
saturation level as a function of the normalization parameter (which is proportional 
to mass accretion rate in the disk),  namely, $\Gamma_{sat}\sim 2.5$. 
We can also make a BH mass estimate for 4U~1630-47 in the  case of  $\Gamma_{sat}\sim 2.5$.  
We choose 
scaling  of the index-normalization previously found in GRO~J1655-40 (see ST09) with that in 4U~1630-47  (see Fig.~\ref{1655_scal}, middle right panel). 

The soft X-ray transient GRO~J1655-40 (Nova Sco 1994) is  a well-studied BHC XRB with the best measured mass  and distance (\cite{HR95}, \cite{Greene01}, see Table 5). Note that the 
correlations  between 4U~1630-47 and  GRO~J1655-40 are self-similar  for the  2000 -- 2001 and  rise 2005 periods respectively.


Thus,  using Eq. ({\ref{geom factor}), we obtain that  $M_{1630}\sim 10~M_{\odot}$ 
($M_{1630}=M_t$) if  
$f_G=1$ (inclinations for both objects are the same, namely $\sim 70$ degrees), 
$s_N=6.00\pm 0.08$, $D_{1630}=10$ kpc, $D_{1655}=3.2$ kpc, $M_{1655}=6.3\pm 0.3~M_{\odot}$
($M_{1655}=M_r$). 


We want to empasize that   scaling  of the correlations  in 4U 1630-47  with that in  XTE~J1550-564, H~1743-322  and  GRO~J1655-40 sources gives  the similar values of BH mass,  about 10 solar masses for  4U 1630-47.

\subsection{Comparison between 4U~1630--47 and GRS~1915-105 
\label{grs1915}} 

BHC binaries 4U~1630--47 and GRS~1915-105 demonstrate a number of similar spectral and timing properties,  however we also would like to point out  some significant differences 
between these objects. 
As a main similar signature 
in these objects, one can consider the same index saturation levels 
($\Gamma_{sat}\sim 3$) and the presence of LFQPOs in the power spectra for both objects, which are classified as the type B 
and C QPOs. Furthermore, the properties of these QPOs are the same in terms of frequency-$S_{rms}$ (the  inverse of $rms$) diagram with 
a very similar tracks for both objects (see bottom left panel of Fig.~\ref{gam_qpo_freq_scal}). 
Despite showing the same saturation levels  of  the index correlation  with the normalization  in GRS~1915+105  and 4U~1630-47  these correlations are not  self-similar, (see the right bottom panel in  Fig. ~\ref{gam_qpo_freq_scal}) and we cannot estimate BH mass in  4U~1630-47 using scaling  with GRS~1915+105. Specifically, 
the  index-normalization diagrams have a different shapes: GRS~1915+105 are characterized by a stepper gradient of the 
index vs  the normalization parameter than that for 4U~1630-47.
In fact, 4U~1630-47 shows a lower level of the index range ($\Gamma_{min,1630}<1.5$) than   that in GRS~1915+105  where the index $\Gamma$  never \ reaches the {\it canonical} LHS  value,  $\Gamma_{min,1915}>1.5$, see e.g. Fender \& Belloni (2004). 

Also we should also point to  
different patterns in the radio/X-ray behavior for these objects. 
Namely, when we compare  radio loud phases [at MJD=50855 -- 50865 for 4U~1630--47, see Fig.~\ref{gam_qpo_freq_scal} 
(top left panel) 
and MJD=50910 -- 51000 for GRS~1915+105,  see top right panel there
 with the almost constant X-ray flux (so called ``$plateau$'' states). 
The peak of radio emission in  GRS~1915  at 15~GHz, 
occurs 
at relatively low X-ray count rate [Soleri et al., 2008])
while the peak of radio emission (4-9 GHz), 
in 4U~1630-47 is  directly related to maximal X-ray count rate 
(Hjellming et al., 1999).



Note also that these objects sometimes demonstrate  properties which are difficult to classify 
in terms of the canonical spectral states (McClintock \& Remillard, 2003). Some {\it anomalous} properties are similar for 
both objects. In particular, during the 1998 outburst rise 4U~1630-47 demonstrated an interesting phase which could be 
considered as a transition between the LHS and the VHS. 
On the other hand, 
the energy spectrum of the source was typically hard, with a well determined exponential high energy cut-off, similar to  that in 
the low state. But the source power spectrum was  a typical for the very high state. Previously, 
Tomsick et al. 
(2005) also revealed a similar {\it anomalous} state in 4U~1630-47  which also revealed a  
disagreement between spectral and timing properties. Specifically, 
in the LHS during 2003 -- 2004 {\it  RXTE} 
observations  4U~1630-47 showed 
spectra 
typical for the  LHS, while the corresponding  power spectra contain  strong 
band limited noise which is rather typical to the HSS. 
Note also that no radio detections during the 2002 -- 2004 period were reported for 4U~1630-47.
All these cases are not consistent with 
the {\it canonical} spectral state sequence 
and indicate  the modified spectral state transition track for 4U~1630-47. 
It should be also  noted that  similar energy spectra related to 
 the aforementioned timing behavior  were  observed at the low-luminosity state of GRS~1915+105 (Trudolyubov et al. 1999a),  during the rise phase in X-ray novae KS~1730-312 (Trudolyubov et al. 1996) and in GS/GRS~1124268 (Ebisawa et al. 1994).


There is also a striking resemblance between the X-ray properties of 4U~1630--47 during the rise of the 1998 outburst,
and GRS~1915+105 in the LHS 
(see Trudolyubov, Churazov \& Gilfanov 1999 and TS09) and  the {\it flaring} 
stage in the VHS [\cite{Tomsick05}]. 
Moreover, radio observations of 4U~1630--47 with the ATCA in 1998 during the brightest phase 
provide a detection of 28\% and 26\% linear polarization at 4.80 and 8.64 GHz, respectively (Hjellming et al., 1999).  The large linear polarization can indicate to  optically thin jet ejections in 4U~1630--47.  The above similarity 
of X-ray spectra of 4U~1630--47 and GRS~1915 %
supports an idea that 4U~1630--47 is also a relativistic jet source.


\section{Conclusions \label{summary}} 

We present  analysis  of the X-ray  spectral and timing  properties observed  from BHC X-ray binary 4U~1630--47 
during  transitions between the hard and soft states. We analyze several 
outbursts  from this source  observed 
with  {\it BeppoSAX}  and 
{\it RXTE} satellites.  
 We apply  the  scaling technique to  correlations of 
the photon index of the Comptonized component versus a low frequency  QPO 
and versus  the mass accretion rate.
We interpret 
the changes of the spectral states 
in 4U~1630--47 in terms of a dynamical evolution of the Compton cloud. 

We show that the X-ray broadband energy spectra of 4U~1630--47 during
all spectral states can be modeled by a combination of a thermal  component, a Comptonized
component, 
and  a redskewed iron-line 
component. 
For our  analysis we utilized the broad spectral coverage and resolution of  $BeppoSAX$ detectors 
from 0.3 to 200 keV along  with frequent monitoring and outstanding time resolution 
 of  $RXTE$ observations in the energy range from 3 to 200 keV. 

In this work we  present  arguments  for the presence of a black hole in 
4U~1630--47 based on the detailed analysis of X-ray spectral evolution in this source. 
We establish that the index monotonically increases during spectral transition from the low-hard
 state to the high-soft state and then finally saturates with the $Comptb$ spectral normalization $N_{com}(\propto \dot M$). 

Moreover, 
the photon index 
saturates  with $\dot M$ at several different levels 
for different outbursts.
In the  case when the photon index $\Gamma$ saturates with $\dot M$ at the value  of 3 
we  also find   a correlations between $\Gamma$ and the centroid of low frequency QPO. 
 ST09 give strong arguments that this index saturation is 
a signature  of converging flow into a BH.
It is important to note that the index monotonic growth  and  saturation at high mass accretion rates during transition from the low-hard to high-soft states 
  has  been recently demonstrated   
for    many other  BH candidates  sources, GX~339-4, GRO~J1655-40, XTE~J1650-500, 4U~1543-47, XTE~J1550-564,  H~1743-322, XTE~J1859+226 (ST09), GRS~1915+105 (TS09), SS~433 (ST10)]. 
 In view of these results  thus we  suggest  the presence of BH in 4U~1630--47.

We also find  that the Comptonized fraction 
 $f$  spans  a wide range from 0.05  to 1, 
which  points
to  variable 
 X-ray illumination of the Compton cloud in  4U~1630--47 during state transitions. 
 Furthermore, we argue that the  changes 
 of $f$ are related to state transitions, 
 which are presumably governed by mass accretion regime.

 Correlations between spectral and timing properties allow us to estimate a BH mass 
 in 4U~1630--47 which is around
10 solar masses and   
the inclination angle $i\lax 70^{\circ}$  
applying the scaling method and using BHC XTE~J~1550-564 
as the reference source.

We acknowledge valuable comments by Chris Shrader during the revision of this Paper.  
We also appreciate the detailed discussion with the referee on the manuscript content.

\newpage

%
%
%
%

\begin{deluxetable}{lcccc}
\tablewidth{0in}
\tabletypesize{\scriptsize}
    \tablecaption{
{\it Beppo}SAX observations of 4U~1630--47 used in analysis.}
    \renewcommand{\arraystretch}{1.2}
\tablehead{
N & Obs. ID& Start time (UT)  & End time (UT) &MJD interval}
\startdata
S1 & 20114001 & 1998 Feb 20 04:30:02 & 1998 Feb 20 16:55:45 &50864.2-50864.7$^1$ \\
S2 & 20114002 & 1998 Feb 24 05:32:00 & 1998 Feb 24 15:00:40 &50868.2-50868.6$^1$ \\
S3 & 20114003 & 1998 Mar 7  12:23:21 & 1998 Mar  8 03:52:21 &50879.5-50880.1$^1$ \\
S4 & 20114004 & 1998 Mar 19 14:52:11 & 1998 Mar 20 04:18:08 &50891.6-50892.3$^1$ \\
S5 & 20114005 & 1998 Mar 26 17:16:48 & 1998 Mar 27 09:59:27 &50898.7-50899.4$^1$ \\
S6 & 70566001 & 1998 Aug  6 18:10:26 & 1998 Aug  7 17:59:20 &51031.7-51031.3$^2$ \\
S7 & 70821005 & 1999 Aug  8 19:26:53 & 1998 Aug 10 17:33:02 &51398.8-51400.1$^2$ \\
      \enddata
   \label{tab:table}
Reference. 
(1) \cite{Ooster98};
(2) \cite{Dieters00} 
\end{deluxetable}

\newpage

%
%

\begin{deluxetable}{lcclcl}
\tablewidth{0in}
\tabletypesize{\scriptsize}
    \tablecaption{
{\it RXTE} observations of 4U~1630--472}
    \renewcommand{\arraystretch}{1.2}
\tablehead{Number of set  & Dates, MJD & RXTE Proposal ID&  Dates UT & Rem. & Ref.}
 \startdata
R1  &    50114        & 00033        & 1996 Feb. 1                &            &      \\
    &    50206-50311  & 10411        & 1996 May 3 -- Aug 16       &            & 1, 2, 3    \\
R2  &    50876-50972  & 30172        & 1998 Mar 4 -- Jun 8        & $Beppo$SAX & 4, 5 \\
    &    50853-50903  & 30178        & 1998 Mar 9 -- Mar 31       & $Beppo$SAX & 4, 6 \\
    &    50855-50873  & 30188        & 1998 Feb 11 -- Mar 1       & $Beppo$SAX & 4, 6 \\
    &    50988-50990  & 30410        & 1998 Jun 24 -- Jun 26      &            &      \\
R3  &    51537-51557  & 40112        & 1999 Dec 25 -- 2000 Jan 14 &            &      \\
    &    51306-51405  & 40418        & 1999 May 8 -- Sep 15       & $Beppo$SAX &      \\
R4  &    51864-51914  & 50120        & 2000 Nov 16 -- 2001 Jan 5  &            &      \\
    &    51917-51978  & 50135        & 2001 Jan 8 -- March 10     &            &      \\
    &    51980-52077  & 60118        & 2001 March 12 -- Jun 17    &            &      \\
R5  &  52519-52654  & 70417        & 2002 Sep 12 -- 2003 Jan 16 &            & 7, 8   \\
    &     52658-52689  & 70113        & 2003 Jan 19 -- Feb 19      &            &   7   \\
    &    52790-53081  & 80117        & 2003 May 31 -- 2004 May 17 &            &   7  \\
      \hline
      \enddata
    \label{tab:par_bbody}
References:
(1) Marshall 1996;    
(2) Levine et al. 1996;  
(3) Kuulkers et al. 1998; 
(4) \cite{Trud01}; 
(5) \cite{Tomsick+Kaaret00};  
(6) \cite{Dieters00};
(7) \cite{Tomsick05}; 
(8) \cite{Homan02}.
\end{deluxetable}

%
%


\newpage
\bigskip
\begin{deluxetable}{cllllllllllll}
\rotate
\tablewidth{0in}
\tabletypesize{\scriptsize}
    \tablecaption{Best-fit parameters of spectral analysis of {\it Beppo}SAX 
observations of 4U~1630--47 in 0.3-200~keV energy range$^{\dagger}$.
Parameter errors correspond to 90\% confidence level.}
    \renewcommand{\arraystretch}{1.2}
 \tablehead
{Observational & MJD, &$\alpha=$  & $\log(A)$ & N$_{Com}^{\dagger\dagger}$ &$T_s$, & $kT_e$, &$kT_{BB}$, & $N_{BB}^{\dagger\dagger}$ &  E$_{line}$,& $N_{line}^{\dagger\dagger}$ &  $N_H$, & $\chi^2_{red}$ (d.o.f.)\\
ID             & day  &$\Gamma-1$ &           &              & keV   & keV   &keV       &                           &   keV       &                             &  cm$^{-2}$ &           }
 \startdata
20114001     &   50864.238 & 1.68(2) & 0.02(1)  & 12.86(4) &  1.49(4) & $>$200  & 0.75(2) & 8.24(7) & 6.51(4) & 0.82(7) & 7.5(1) & 1.05(127) \\
20114002     &   50868.254 & 1.62(5) & -0.18(3) & 11.83(3) &  1.48(3) & $>$200  & 0.74(2) & 8.76(6) & 6.42(4) & 0.81(8) & 7.8(1) & 1.16(127) \\
20114003     &   50879.516 & 1.57(6) & -0.15(5) & 11.03(4) &  1.64(5) & $>$200  & 0.74(2) & 13.64(5)& 5.77(8) & 0.65(4) & 7.2(3) & 1.08(127) \\
20114004     &   50891.620 & 1.00(1) & -0.04(2) & 1.80(2)  &  1.46(6) & 50(2)   & 0.75(3) & 3.98(4) & 7.19(3) & 0.16(2) & 6.6(2) & 1.10(127) \\
20114005     &   50898.720 & 1.03(3) & 0.20(5)  & 1.44(2)  &  1.01(7) & 58(1)   & 0.62(4) & 1.96(4) & 7.08(6) & 0.16(7) & 6.6(2) & 1.18(127) \\
70566001     &   51031.516 & 0.97(5) & 2.0$^{\dagger\dagger\dagger}$  & 0.59(3) & 1.6(1)  & 140(20) & 0.75(5) & 0.49(4) & 5.26(4) & 0.10(1) & 6.9(3) & 1.15(128) \\
70821005     &   51398.857 & 1.01(2) & 0.02(1)  & 2.20(3)  &  1.6(2)  & 160(50) & 0.72(9) & 0.07(2) & 5.38(5) & 0.10(2) & 6.9(1) & 0.92(127) \\
      \enddata
    \label{tab:fit_table}
$^\dagger$ The spectral model is  $wabs*(blackbody + Comptb + Laor)$,
$^{\dagger\dagger}$ normalization parameters of $blackbody$ and $Comptb$ components are in units of 
$L_{37}/d^2_{10}$ $erg/s/kpc^2$, where $L_{37}$ is the source luminosity in units of 10$^{37}$ erg/s, 
$d^2_{10}$ is the distance to the source in units of 10 kpc 
and $Laor$ component is in units of $10^{-2}\times total~~photons$ $cm^{-2}s^{-1}$ in line, 
$^{\dagger\dagger\dagger}$ when parameter $\log(A)\gg1$, it is fixed to a value 2.0 (see comments in the text). 
\end{deluxetable}


\newpage
\bigskip
\begin{deluxetable}{llllllllllllll}
\rotate
\tablewidth{0in}
\tabletypesize{\scriptsize}
    \tablecaption{Best-fit parameters of spectral analysis for 1998 (``$R2$'' set) with PCA\&HEXTE/{\it RXTE} 
observations of 4U~1630--47 in 3 -- 200~keV energy range$^{\dagger}$.
Parameter errors correspond to 90\% confidence level.}
    \renewcommand{\arraystretch}{1.2}
 \tablehead
{Observational & MJD, & $\alpha=$  & $kT_e,$ & $\log(A)$$^{\dagger\dagger}$ & N$_{com}^{\dagger\dagger\dagger}$ & $T_s$,& $N_{Bbody}^{\dagger\dagger\dagger}$ & E$_{line}$,&   $N_{line}^{\dagger\dagger\dagger}$ &  $\chi^2_{red}$ (d.o.f.)& F$_1$/F$_2^{\dagger\dagger\dagger\dagger}$ \\
ID             & day  & $\Gamma-1$ & keV    &                           &                                       & keV&                                     &  keV       &                                      &                         &                                                & }
 \startdata 
30178-01-01-00 & 50853.057 & 1.06(1) &  48(8)     & 0.92(8) & 2.18(6) & 1.00(6) & 1.06(9) &  6.10(9) & 0.63(9) & 1.41(93) & 1.97/3.17 \\
30178-01-02-00 & 50853.655 & 1.47(9) &  60$\pm$10 & 2.00$^{\dagger\dagger}$& 4.2(1)  & 1.9(2)  & 3.0(1)  &  5.9(1)  & 0.19(7) & 0.73(94) & 2.61/3.82 \\
30178-02-01-00 & 50855.046 & 1.23(1) &  50$\pm$10 & 2.00$^{\dagger\dagger}$& 8.08(9) & 1.03(9) & 1.14(5) &  6.40(6) & 0.13(6) & 1.43(94) & 5.14/5.96 \\
30188-02-01-00 & 50855.428 & 1.28(2) &  43(3)     & 2.00$^{\dagger\dagger}$& 8.04(5) & 1.13(8) & 3.56(9) &  6.41(3) & 0.16(6) & 1.25(94) & 5.53/6.22 \\
30178-01-03-00 & 50855.841 & 1.54(4) &  70$\pm$20 & 1.01(9) & 7.5(1)  & 1.9(1)  & 7.89(9) &  5.61(5) & 0.03(8) & 0.90(93) & 6.19/6.10 \\
30188-02-02-00 & 50856.115 & 1.30(2) &  52(3)     & 2.00$^{\dagger\dagger}$& 8.5(1)  & 1.13(9) & 4.2(1)  &  6.40(6) & 0.18(6) & 0.86(94) & 5.95/6.24 \\
30188-02-04-00 & 50856.502 & 1.36(2) &  51(7)     & 2.00$^{\dagger\dagger}$& 9.2(1)  & 1.13(8) & 4.6(1)  &  6.39(2) & 0.19(4) & 1.32(94) & 6.55/6.39 \\
30178-01-04-00 & 50856.627 & 1.58(4) &  50$\pm$10 & 1.62(9) & 7.5(1)  & 1.9(2)  & 7.98(9) &  5.67(3) & 0.06(9) & 1.27(93) & 6.42/6.16 \\
30188-02-03-00 & 50856.651 & 1.37(2) &  54(4)     & 2.00$^{\dagger\dagger}$& 8.9(1)  & 1.13(9) & 4.6(1)  &  6.42(6) & 0.21(9) & 1.18(94) & 6.38/6.25 \\
30178-02-01-01 & 50856.869 & 1.48(2) &  54(3)     & 2.00$^{\dagger\dagger}$& 10.0(1) & 1.02(9) & 3.29(5) &  6.41(7) & 0.19(6) & 1.26(94) & 6.54/6.38 \\
30178-01-05-00 & 50857.116 & 1.64(6) &  90$\pm$20 & 2.00$^{\dagger\dagger}$& 7.5(1)  & 1.9(1)  & 8.64(1) &  5.70(2) & 0.08(2) & 1.24(94) & 6.84/6.04 \\
30188-02-05-00 & 50857.709 & 1.58(2) & 134(7)     & 0.92(1) & 9.16(9) & 1.12(9) & 4.99(8) &  6.42(3) & 0.19(6) & 0.89(93) & 6.57/5.04 \\
30178-02-02-00 & 50857.793 & 1.54(3) & 105(5)     & 2.00$^{\dagger\dagger}$& 10.1(1) & 1.03(8) & 4.03(8) &  6.41(5) & 0.14(7) & 0.81(94) & 6.85/5.35 \\
30188-02-06-00 & 50858.045 & 1.46(2) & 102(6)     & 2.00$^{\dagger\dagger}$& 9.57(9) & 1.13(9) & 4.69(9) &  6.40(6) & 0.12(7) & 0.73(94) & 6.72/5.76 \\
30178-01-06-00 & 50858.694 & 1.62(5) &  70$\pm$20 & 2.00$^{\dagger\dagger}$& 7.6(1)  & 1.9(1)  & 8.49(1) &  5.73(3) & 0.11(9) & 0.71(94) & 6.92/5.78 \\
30188-02-07-00 & 50858.715 & 1.49(2) & 105(8)     & 2.00$^{\dagger\dagger}$& 9.83(9) & 1.14(8) & 4.86(9) &  6.44(8) & 0.16(6) & 0.80(94) & 6.93/5.77 \\
30178-02-02-01 & 50858.771 & 1.53(1) & 113(6)     & 2.00$^{\dagger\dagger}$& 9.97(7) & 1.03(9) & 4.07(4) &  6.40(5) & 0.18(5) & 1.39(94) & 6.79/5.65 \\
30178-01-07-00 & 50859.836 & 1.72(5) &  50$\pm$20 & 2.00$^{\dagger\dagger}$& 7.6(1)  & 1.85(8) & 9.19(8) &  5.89(6) & 0.08(9) & 0.73(94) & 7.28/5.35 \\
30188-02-08-00 & 50860.117 & 1.55(3) & 118(3)     & 0.79(9) & 10.7(1) & 1.13(9) & 4.51(9) &  6.40(2) & 0.19(6) & 0.80(93) & 7.16/5.13 \\
30188-02-09-00 & 50860.562 & 1.60(4) & 148(7)     & 2.00$^{\dagger\dagger}$& 9.9(1)  & 1.12(7) & 5.6(1)  &  6.40(5) & 0.19(8) & 0.92(94) & 7.30/5.69 \\
30178-01-08-00 & 50860.722 & 1.70(4) & 160$\pm$20 & 2.00$^{\dagger\dagger}$& 7.7(2)  & 1.9(1)  & 9.46(9) &  5.50(6) & 0.11(7) & 0.82(94) & 7.40/5.97 \\
30178-01-09-00 & 50861.693 & 1.70(4) & 130(8)     & 2.00$^{\dagger\dagger}$& 9.26(2) & 1.67(4) & 9.26(2) &  6.41(3) & 0.19(4) & 0.94(94) & 7.37/6.02 \\
30188-02-10-00 & 50861.723 & 1.52(2) & 125(2)     & 2.00$^{\dagger\dagger}$& 9.98(9) & 1.13(8) & 5.5(1)  &  6.40(2) & 0.19(6) & 0.93(94) & 7.36/5.98 \\
30178-01-10-00 & 50862.653 & 1.71(4) & 165(5)     & 2.00$^{\dagger\dagger}$& 7.85(2) & 2.0(2)  & 9.52(2) &  6.41(6) & 0.14(6) & 0.95(94) & 7.50/5.88 \\
30188-02-11-00 & 50862.704 & 1.51(2) & 125(3)     & 2.00$^{\dagger\dagger}$& 10.1(1) & 1.12(8) & 5.7(1)  &  6.41(4) & 0.11(5) & 0.94(94) & 7.41/6.04 \\
30178-02-03-00 & 50862.765 & 1.54(1) &  96(4)     & 2.00$^{\dagger\dagger}$& 10.95(9)& 1.07(9) & 5.05(9) &  6.40(5) & 0.13(3) & 1.09(94) & 7.65/6.22 \\
30178-01-11-00 & 50863.695 & 1.71(4) & 127(7)     & 0.64(9) & 9.50(1) & 1.7(2)  & 12.38(1)&  6.43(4) & 0.18(5) & 0.93(93) & 9.40/5.44 \\
30188-02-12-00 & 50863.712 & 1.87(3) & 180$\pm$20 & 0.66(8) & 14.7(1) & 1.12(8) & 5.5(1)  &  6.41(4) & 0.21(8) & 0.90(93) & 9.46/5.51 \\
30188-02-13-00 & 50864.185 & 1.87(4) & 125(4)     & 0.91(9) & 15.97(1)& 1.13(9) & 5.8(1)  &  6.47(3) & 0.19(5) & 0.87(93) &10.30/7.24 \\
30188-02-14-00 & 50864.320 & 1.92(3) & $>$200     & 1.3(1)  & 12.9(1) & 1.14(7) & 5.91(8) &  6.45(5) & 0.15(6) & 1.06(93) & 8.46/6.00 \\
30178-01-12-00 & 50864.630 & 1.70(2) & 130(4)     & 0.28(3) & 9.22(6) & 1.93(8) & 13.99(9)&  6.46(2) & 0.17(8) & 1.09(93) & 9.54/5.43 \\
30188-02-15-00 & 50865.046 & 1.73(4) & 125(7)     & 0.9(1)  & 15.98(7)& 1.13(9) & 5.7(1)  &  6.40(6) & 0.19(6) & 0.72(93) &10.24/6.82 \\
30188-02-16-00 & 50865.320 & 1.69(3) & 161(9)     & 1.7(1)  & 11.48(9)& 1.12(6) & 6.1(1)  &  6.44(4) & 0.13(3) & 0.94(93) & 8.32/5.90 \\
30188-02-17-00 & 50866.639 & 1.70(4) & 172(3)     & 0.25(4) & 18.0(1) & 1.12(7) & 6.0(1)  &  6.48(5) & 0.19(6) & 1.21(93) & 9.70/5.76 \\
30178-01-13-00 & 50867.497 & 1.72(3) & 200$\pm$40 & 0.27(7) & 8.34(9) & 1.92(9) & 14.36(9)&  6.41(4) & 0.14(9) & 0.86(93) & 9.64/5.44 \\
30188-02-18-00 & 50868.568 & 1.88(2) & $>$200     & 0.28(4) & 15.6(1) & 1.14(2) & 3.0(1)  &  6.40(2) & 0.19(6) & 1.15(93) & 9.11/4.66 \\
30188-02-19-00 & 50869.650 & 1.78(4) & 200$\pm$20 & 0.05(2) & 17.7(1) & 1.12(7) & 6.04(9) &  6.42(3) & 0.17(1) & 1.23(93) & 9.13/4.79 \\
30188-02-20-00 & 50870.181 & 1.44(1) & 70$\pm$20  & 2.00$^{\dagger\dagger}$& 5.7(1)  & 1.13(5) & 7.7(1)  &  6.40(4) & 0.19(6) & 1.13(94) & 5.43/3.97 \\
30178-01-14-00 & 50870.978 & 1.70(2) & 150$\pm$30 & 2.00$^{\dagger\dagger}$& 4.93(9) & 1.98(3) & 11.57(9)&  6.41(5) & 0.18(7) & 0.76(94) & 6.15/3.77 \\
30188-02-21-00 & 50871.257 & 1.53(1) & 100$\pm$30 & 2.00$^{\dagger\dagger}$& 5.9(1)  & 1.13(9) & 8.5(1)  &  6.43(4) & 0.19(6) & 0.93(94) & 6.22/3.64 \\
30188-02-21-01 & 50871.410 & 1.49(2) & 100$\pm$10 & 2.00$^{\dagger\dagger}$& 5.97(9) & 1.11(7) & 7.9(1)  &  6.40(4) & 0.12(5) & 0.91(94) & 6.13/3.96 \\
30188-02-22-00 & 50872.841 & 1.51(3) & 103(7)     & 2.00$^{\dagger\dagger}$& 4.87(4) & 1.99(7) & 8.9(1)  &  9.47(1) & 0.19(6) & 0.94(94) & 5.84/4.10 \\
30188-02-23-00 & 50873.138 & 1.52(2) & 126(5)     & 2.00$^{\dagger\dagger}$& 4.61(4) & 1.47(9) & 10.0(1) &  8.46(7) & 0.11(5) & 0.74(94) & 6.15/3.35 \\
30178-01-15-00 & 50874.516 & 1.54(8) & 100$\pm$20 & 0.3(1)  & 8.92(1) & 1.40(3) & 7.77(9) &  6.40(4) & 0.08(6) & 1.19(93) & 6.97/4.19 \\
30172-01-01-01 & 50876.184 & 1.34(4) &  56(5)     & 0.49(9) & 7.0(1)  & 0.99(6) & 4.1(2)  &  6.41(1) & 0.06(2) & 0.63(94) & 5.22/3.68 \\
30172-01-01-02 & 50877.186 & 1.29(2) &  43(2)     & 0.10(2) & 9.59(3) & 1.05(4) & 6.74(9) &  6.43(2) & 0.03(1) & 0.82(94) & 5.13/3.98 \\
30172-01-01-03 & 50880.378 & 1.61(3) & 120$\pm$20 & 0.27(7) & 12.6(2) & 1.11(6) & 5.51(8) &  6.40(1) & 0.04(8) & 0.66(94) & 6.58/4.72 \\
30172-01-01-04 & 50881.378 & 1.69(2) &  50$\pm$10 & 0.33(8) & 14.1(2) & 1.05(8) & 1.8(1)  &  6.49(4) & 0.03(4) & 1.12(93) & 7.33/4.78 \\
30178-01-16-00 & 50881.580 & 1.54(8) & 130$\pm$10 & 0.3(1)  & 8.92(1) & 1.4(1)  & 7.77(9) &  6.40(6) & 0.19(6) & 1.11(93) & 6.96/4.19 \\
30172-01-01-00 & 50883.847 & 1.51(7) & 60$\pm$30  & 0.01(1) & 4.8(1)  & 1.01(6) & 5.0(1)  &  6.44(1) & 0.02(1) & 1.01(94) & 4.24/1.70 \\
30178-01-17-00 & 50884.981 & 1.45(3) & 100$\pm$20 & -0.29(8)& 7.89(5) & 1.00(3) & 3.81(9) &  6.40(9) & 0.19(6) & 0.95(93) & 3.88/1.76 \\
30172-01-02-01 & 50885.146 & 1.20(1) &  60(8)     & -0.18(2)& 6.14(4) & 1.10(8) & 9.84(5) &  6.41(1) & 0.04(1) & 0.86(93) & 3.55/2.14 \\
30172-01-02-00 & 50885.805 & 1.27(2) &  55(3)     & -0.10(2)& 5.79(7) & 1.10(7) & 10.0(1) &  6.40(3) & 0.02(4) & 0.98(93) & 3.44/2.06 \\
30172-01-03-00 & 50887.714 & 1.22(1) &  43(4)     & -0.07(3)& 4.6(1)  & 1.07(5) & 4.41(9) &  6.43(2) & 0.03(2) & 1.04(93) & 3.08/1.75 \\
30178-01-18-00 & 50889.714 & 1.30(3) &  55(4)     & -0.23(5)& 4.74(6) & 1.00(3) & 3.48(7) &  6.44(6) & 0.19(6) & 1.07(93) & 2.90/1.34 \\
30172-01-05-00 & 50891.646 & 1.13(1) &  43(2)     & -0.07(4)& 3.21(8) & 1.07(5) & 4.02(8) &  6.41(1) & 0.03(1) & 1.35(93) & 2.27/1.35 \\
30172-01-04-00 & 50892.584 & 1.15(2) &  41(5)     & -0.11(4)& 3.15(8) & 1.07(6) & 4.01(8) &  6.40(2) & 0.04(3) & 1.35(93) & 2.22/1.28 \\
30172-01-06-00 & 50893.719 & 1.17(3) &  50(7)     & -0.36(9)& 4.04(5) & 0.99(4) & 2.91(7) &  6.45(6) & 0.02(1) & 0.89(93) & 2.09/1.09 \\
30172-01-07-00 & 50895.651 & 1.11(2) &  45(2)     & 0.39(6) & 1.86(8) & 1.23(9) & 2.67(9) &  6.50(1) & 0.07(2) & 1.14(93) & 1.62/1.35 \\
30178-01-19-00 & 50897.802 & 1.15(2) &  48(6)     & 0.36(4) & 2.44(5) & 1.00(3) & 2.70(8) &  6.42(3) & 0.19(6) & 1.05(93) & 1.40/1.15 \\
30172-01-08-00 & 50899.001 & 0.96(5) &  56(6)     & 0.03(2) & 1.82(9) & 1.30(9) & 3.38(9) &  6.50(1) & 0.06(1) & 1.22(93) & 1.36/1.10 \\
30172-01-08-02 & 50900.675 & 1.01(3) &  50(7)     & 0.18(6) & 1.6(1)  & 1.31(8) & 2.9(1)  &  5.74(6) & 0.07(2) & 0.85(93) & 1.15/1.04 \\
30410-02-06-00 & 50900.747 & 1.00(4) &  90$\pm$20 & 2.00$^{\dagger\dagger}$& 3.14(1) & 1.27(9) & 3.0(1)  &  5.9(1)  & 0.01(9) & 1.17(94) & 0.18/0.34 \\
30178-01-20-00 & 50903.139 & 1.08(3) &  51(5)     & -0.01(1)& 1.90(5) & 1.00(3) & 2.38(9) &  6.40(5) & 0.19(6) & 1.29(93) & 1.08/0.93 \\
30172-01-08-03 & 50904.209 & 1.17(4) &  50(8)     & -0.12(4)& 1.87(2) & 1.01(8) & 3.1(1)  &  6.34(6) & 0.07(5) & 1.01(93) & 1.02/0.85 \\
30172-01-09-00 & 50904.525 & 0.91(3) &  51(1)     & 0.08(2) & 1.5(1)  & 1.29(5) & 3.58(2) &  6.36(7) & 0.04(6) & 1.05(93) & 1.00/0.96 \\
30172-01-10-00 & 50906.513 & 0.99(1) &  50(3)     & 0.2(1)  & 1.6(1)  & 1.20(4) & 3.67(9) &  6.56(6) & 0.08(3) & 1.39(93) & 0.92/1.08 \\
30172-01-11-00 & 50907.511 & 1.00(1) &  49(5)     & 0.17(8) & 1.57(9) & 1.24(3) & 3.76(5) &  6.57(3) & 0.11(6) & 1.20(93) & 0.89/1.04 \\
30172-01-12-00 & 50909.410 & 1.00(2) &  50(3)     & 0.11(4) & 1.59(7) & 1.00(5) & 3.88(2) &  5.9(2)  & 0.07(5) & 0.91(93) & 0.89/0.98 \\
30172-01-13-00 & 50911.309 & 1.06(3) &  44(4)     & -0.07(8)& 1.6(1)  & 1.09(3) & 3.91(9) &  6.5(1)  & 0.09(6) & 0.73(93) & 0.91/0.73 \\
30172-01-14-00 & 50913.311 & 1.00(1) &  51(2)     & -0.22(5)& 1.54(9) & 1.11(2) & 3.6(1)  &  5.98(9) & 0.03(7) & 0.69(93) & 0.88/0.61 \\
30172-01-08-04 & 50921.580 & 0.93(7) &  50(6)     & 0.10(8) & 1.74(5) & 1.01(9) & 3.49(9) &  6.34(6) & 0.05(6) & 1.14(93) & 0.96/1.15 \\
30172-01-15-00 & 50923.598 & 1.05(2) &  55(3)     & -0.13(4)& 1.91(9) & 1.08(2) & 3.61(6) &  7.6(2)  & 0.01(3) & 0.73(93) & 0.96/0.82 \\
30172-01-16-00 & 50924.744 & 1.03(2) &  47(2)     & -0.09(4)& 2.1(1)  & 1.09(3) & 3.24(6) &  6.8(1)  & 0.06(8) & 1.00(93) & 1.01/0.97 \\
30172-01-17-00 & 50925.857 & 1.03(4) &  50$\pm$10 & 0.13(7) & 1.7(1)  & 1.06(8) & 2.38(9) &  6.82(8) & 0.02(6) & 1.03(93) & 1.05/1.03 \\
30172-01-17-01 & 50926.653 & 0.89(3) &  60$\pm$10 & 0.06(3) & 1.9(1)  & 1.07(7) & 2.4(1)  &  6.79(3) & 0.01(5) & 1.00(93) & 1.10/1.19 \\
30172-01-17-02 & 50927.722 & 1.1(1)  &  72$\pm$20 & 0.03(3) & 1.82(9) & 1.05(9) & 2.6(1)  &  6.80(4) & 0.06(3) & 0.99(93) & 1.17/0.95 \\
30172-01-17-03 & 50928.591 & 1.22(6) &  90$\pm$20 & 0.05(4) & 1.98(9) & 1.01(7) & 2.48(9) &  6.83(8) & 0.04(6) & 1.10(93) & 1.19/0.90 \\
30172-01-17-04 & 50929.657 & 1.10(6) & 100$\pm$20 & -0.04(2)& 2.0(1)  & 1.07(6) & 2.64(7) &  6.84(5) & 0.05(5) & 0.90(93) & 1.22/0.93 \\
30172-01-17-05 & 50930.657 & 1.07(6) &  90$\pm$20 & -0.13(7)& 2.2(1)  & 1.05(7) & 2.82(9) &  6.80(8) & 0.03(6) & 0.89(93) & 1.28/0.92 \\
30172-01-17-06 & 50931.859 & 1.09(7) &  50$\pm$10 & 0.09(7) & 1.7(1)  & 1.05(7) & 2.05(8) &  6.81(8) & 0.07(8) & 1.03(93) & 1.13/0.94 \\
30172-01-17-07 & 50932.793 & 1.10(5) &  90$\pm$10 & -0.02(9)& 2.34(9) & 1.02(8) & 2.64(9) &  6.82(3) & 0.08(7) & 1.05(93) & 1.35/1.11 \\
30172-01-18-00 & 50935.127 & 1.08(3) & 110$\pm$20 & -0.16(8)& 2.6(1)  & 0.99(5) & 2.69(7) &  6.80(2) & 0.11(6) & 0.78(93) & 1.42/1.04 \\
30172-01-18-01 & 50936.194 & 1.15(5) &  90$\pm$10 & -0.25(9)& 2.6(1)  & 0.97(6) & 2.65(6) &  6.80(8) & 0.09(9) & 0.84(93) & 1.39/0.85 \\
30172-01-18-02 & 50937.622 & 1.06(7) &  80$\pm$10 & -0.14(8)& 2.25(8) & 1.08(7) & 3.16(9) &  6.82(6) & 0.04(6) & 0.88(93) & 1.41/0.96 \\
30172-01-18-03 & 50939.061 & 1.00(4) &  90$\pm$10 & -0.05(6)& 2.0(1)  & 1.14(6) & 3.05(9) &  6.85(8) & 0.07(2) & 0.84(93) & 1.32/1.01 \\
30172-01-18-04 & 50942.015 & 0.81(7) &  50$\pm$20 & -0.08(7)& 1.38(9) & 1.2(1)  & 2.6(1)  &  6.86(7) & 0.03(6) & 1.05(93) & 1.09/0.81 \\
30172-01-18-05 & 50945.860 & 0.9(1)  &  90$\pm$10 & 0.14(9) & 1.2(1)  & 1.3(1)  & 5.98(8) &  6.60(9) & 0.02(5) & 1.33(93) & 0.93/0.76 \\
30172-01-18-06 & 50945.861 & 1.01(5) &  40$\pm$10 & 0.2(1)  & 0.99(2) & 1.21(6) & 2.46(9) &  6.84(8) & 0.03(6) & 1.32(93) & 0.71/0.67 \\
30172-01-18-07 & 50949.732 & 0.89(7) &  50$\pm$10 & 0.24(7) & 0.92(2) & 1.39(5) & 2.24(6) &  5.99(9) & 0.06(6) & 1.01(93) & 0.64/0.69 \\
30172-01-18-08 & 50951.661 & 0.93(2) &  60$\pm$10 & 2.00$^{\dagger\dagger}$& 0.91(4) & 1.00(6) & 0.91(6) &  6.21(7) & 0.03(4) & 0.87(94) & 0.59/1.05 \\
30172-01-18-09 & 50952.482 & 0.86(3) &  40$\pm$10 & 2.00$^{\dagger\dagger}$& 0.83(3) & 1.00(7) & 0.78(5) &  6.18(7) & 0.33(6) & 1.46(94) & 0.52/1.03 \\
30172-01-18-10 & 50953.983 & 0.77(2) &  50$\pm$10 & 2.00$^{\dagger\dagger}$& 0.62(2) & 1.00(3) & 0.62(4) &  6.03(6) & 0.21(5) & 1.12(94) & 0.42/0.91 \\
30172-01-18-11 & 50956.949 & 0.87(4) &  50$\pm$10 & 2.00$^{\dagger\dagger}$& 0.45(3) & 1.00(6) & 0.35(5) &  6.1(1)  & 0.20(6) & 1.43(94) & 0.28/0.55 \\
30172-01-18-12 & 50958.950 & 0.62(3) &  49(6)     & 2.00$^{\dagger\dagger}$& 0.47(4) & 1.00(7) & 0.41(5) &  5.86(6) & 0.27(9) & 1.00(94) & 0.28/0.72 \\
30172-01-18-13 & 50961.087 & 0.71(6) &  50(8)     & 2.00$^{\dagger\dagger}$& 0.45(5) & 1.00(1) & 0.33(9) &  6.2(1)  & 0.18(2) & 0.82(94) & 0.27/0.65 \\
30172-01-18-14 & 50962.949 & 0.64(3) &  46(6)     & 2.00$^{\dagger\dagger}$& 0.59(3) & 1.00(8) & 0.24(6) &  5.9(1)  & 0.18(6) & 1.42(94) & 0.32/0.89 \\
30172-01-18-15 & 50965.948 & 0.23(3) & 148(9)     & 2.00$^{\dagger\dagger}$& 0.31(4) & 1.00(6) & 0.31(5) &  5.84(9) & 0.11(5) & 0.83(94) & 0.28/0.69 \\
30172-01-18-16 & 50968.261 & 0.49(3) &  50(3)     & 2.00$^{\dagger\dagger}$& 0.47(3) & 1.00(7) & 0.30(5) &  5.84(9) & 0.18(7) & 1.21(94) & 0.27/0.68 \\
30172-01-18-17 & 50970.387 & 0.52(4) &  50(5)     & 2.00$^{\dagger\dagger}$& 0.42(4) & 1.00(4) & 0.28(6) &  5.9(1)  & 0.17(5) & 1.17(94) & 0.24/0.65 \\
30172-01-18-18 & 50972.126 & 0.59(2) &  54(7)     & 2.00$^{\dagger\dagger}$& 0.52(2) & 1.00(6) & 0.19(5) &  6.04(8) & 0.18(4) & 1.39(94) & 0.28/0.82 \\
30410-02-05-00 & 50988.772 & 1.19(5) &  79(2)     & 2.00$^{\dagger\dagger}$& 4.61(4) & 1.47(9) & 10.0(1) &  8.46(9) & 0.01(9) & 0.18(94) & 0.18/0.34 \\
30410-02-07-00 & 50990.885 & 0.35(3) &  85(8)     & 2.00$^{\dagger\dagger}$& 0.26(9) & 0.82(9) & 0.18(9) &  7.14(8) & 0.19(6) & 0.86(94) & 0.18/0.41 \\
      \enddata
    \label{tab:fit_table}
$^\dagger$ The spectral model is  $wabs*(blackbody + Comptb + Laor)$, where $N_H$ is fixed at a value  7.7$\times 10^{22}$ cm$^{-2}$ \citep{Dieters00}; 
$kT_{BB}$ are fixed at 0.7 keV (see comments in the text); 
$^{\dagger\dagger}$ when parameter $\log(A)\gg2$, this parameter is fixed at 2.0 (see comments in the text), 
$^{\dagger\dagger\dagger}$ normalization parameters of $blackbody$ and $Comptb$ components are in units of 
$10^{-2}\times L_{37}/d^2_{10}$ $erg/s/kpc^2$, where $L_{37}$ is the source luminosity in units of 10$^{37}$ erg/s, 
$d^2_{10}$ is the distance to the source in units of 10 kpc 
and $Laor$ component is in units of $10^{-2}\times total~~photons$ $cm^{-2}s^{-1}$ in line, 
$^{\dagger\dagger\dagger\dagger}$spectral fluxes (F$_1$/F$_2$) in units of $\times 10^{-9}$ ergs/s/cm$^2$ for  (3 -- 10) and (10 -- 50) keV energy ranges respectively.  
\end{deluxetable}
~~~~

%
%
\begin{deluxetable}{lcccccc}
\tablewidth{0in}
\tabletypesize{\scriptsize}
    \tablecaption{BH masses, distances and inclination angle determination} 
    \renewcommand{\arraystretch}{1.2}
\tablehead{
     Source   & M$^a_{dyn}$ (M$_{\odot})$ & i$_{orb}^a$ (deg) & d$^a$ (kpc)  & i$_{scal}^a$ (deg) & M$_{scal}$ (M$_{\odot}$)  &  Refs} 
 \startdata
XTE~J1550-564  &   9.5$\pm$1.1 &  72$\pm$5    &   $\sim$6           &        ...     & 10.7$\pm$1.5$^c$ &  1, 2, 3 \\
GRO~J1655-40   &   6.3$\pm$0.3 &  70$\pm$1    &   3.2$\pm$0.2       &        ...     & ...              &  4, 5 \\
H~1743-322     &   $\sim$11    &  $\sim$70    &   $\sim$10          &        ...     & 13.3$\pm$3.2$^c$ &  6  \\
GRS~1915+105   &   14$\pm$4    &  $\sim$60    &   12.1$\pm$0.8      &        ...     & 15.6$\pm$1.5$^d$ &  4, 7  \\
4U~1630--47    &        ...    &     ...      &         $\sim$10    &        $\leq$70  & 9.5$\pm$1.1    & this work \\
      \hline
      \enddata
    \label{tab:par_scal}
$^a$ Dynamically determined BH mass and system inclination angle, $^b$ Source distance found in literature, 
$^c$ Scaling value found by ST09, 
$^d$ Scaling value found by ST07.\\  
References: 
(1) Orosz et al. 2002; 
(2) S$\grave a$nchez-Fern$\grave a$ndez et al. 1999; 
(3) Sobczak et al. 1999; 
(4) Green et al. 2001; (5) Hjellming \& Rupen 1995
(6) McClintock et al. 2007;  
(7) ST09.
\end{deluxetable}
~~~~~~~~~~~~~~~~~~~~~~~~~~~~~~~~~~~~~~~~~~~~~~~~~~~~~~
~~~~~~~~~~~~~~~~~~~~~~~~~~~~~~~~~~~~~~~~~~~~~~~~~~~~~
~~~~~~~~~~~~~~~~~~~~~~~~~~~~~~~~~~~~~~~~~~~~~~~~~~~~~~
~~~~~~~~~~~~~~~~~~~~~~~~~~~~~~~~~~~~~~~~~~~~~~~~~~~~~

          
%
%

\newpage

\begin{figure}[ptbptbptb]
\includegraphics[scale=0.95, angle=0]{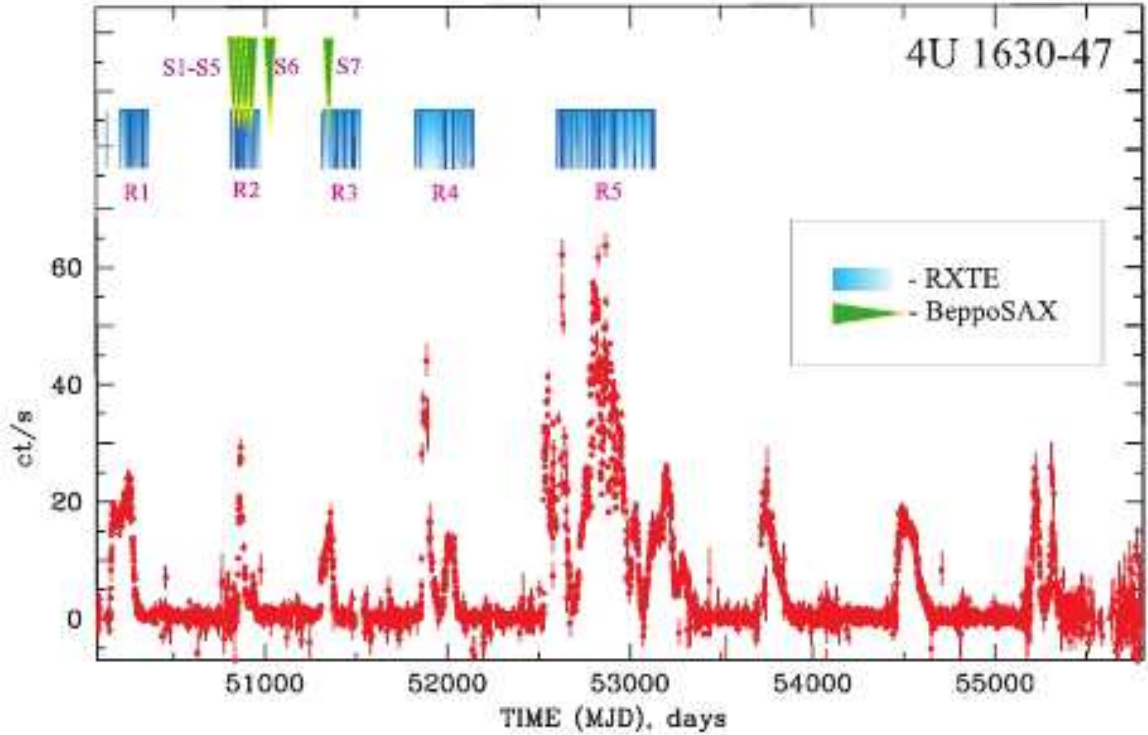}
\caption{
 Evolution of ASM/{\it RXTE} count rate 
during  1996 -- 2011 observations of 4U~1630--47. 
{\it Blue} vertical strips (at {\it top of the panel}) indicate temporal distribution of the {\it RXTE} data 
of pointed observations used in our analysis, whereas {\it bright blue} rectangles indicate
the {\it RXTE} data sets listed in Table 2, 
and {\it green} triangles show {\it Beppo}SAX NFI data, listed in Table 1.
}
\label{asm_1630}
\end{figure}

%
%

\newpage
\begin{figure}[ptbptbptb]
\includegraphics[scale=0.9, angle=0]{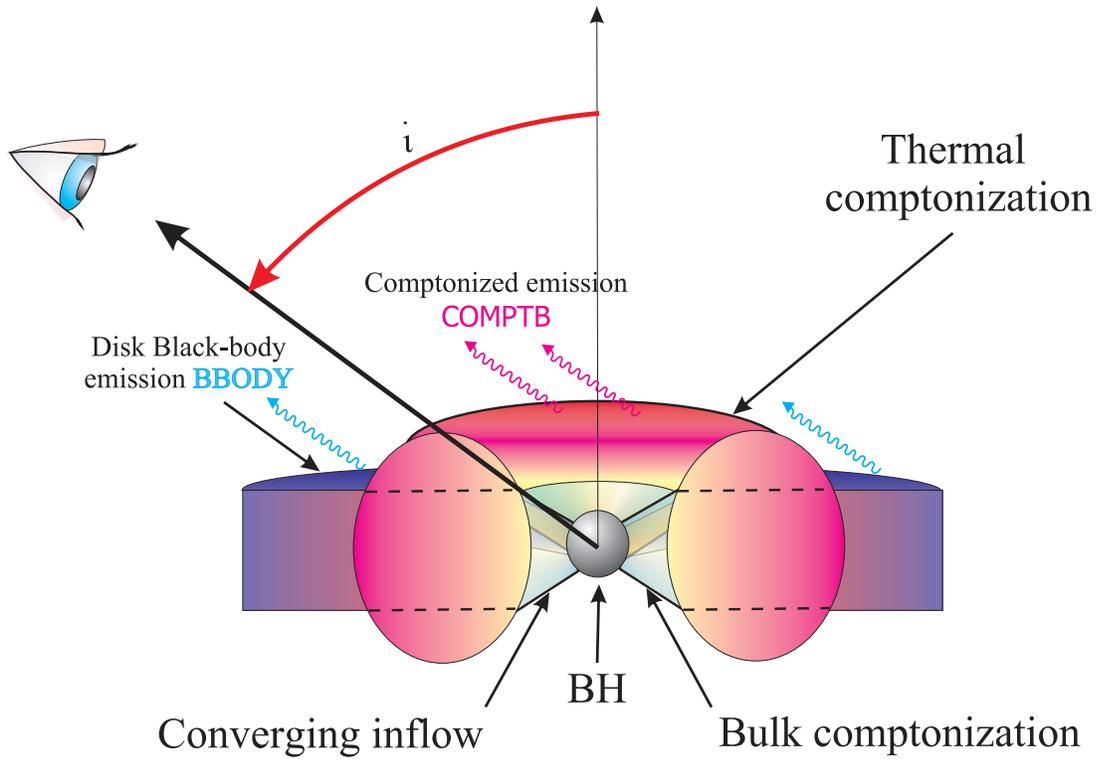}
\caption{A suggested  geometry of the system.   Disk  
soft photons are upscattered (Comptonized) off  relatively hot plasma of the transition layer.  Some fraction of these photons are  directly seen  by the Earth observer.  Blue and pink  photon trajectories correspond to soft (disk) and hard (Comptonized)  photons respectively.
}
\label{geometry}
\end{figure}

%
%

\newpage 
\begin{figure}[ptbptbptb]
\includegraphics[scale=0.87,angle=0]{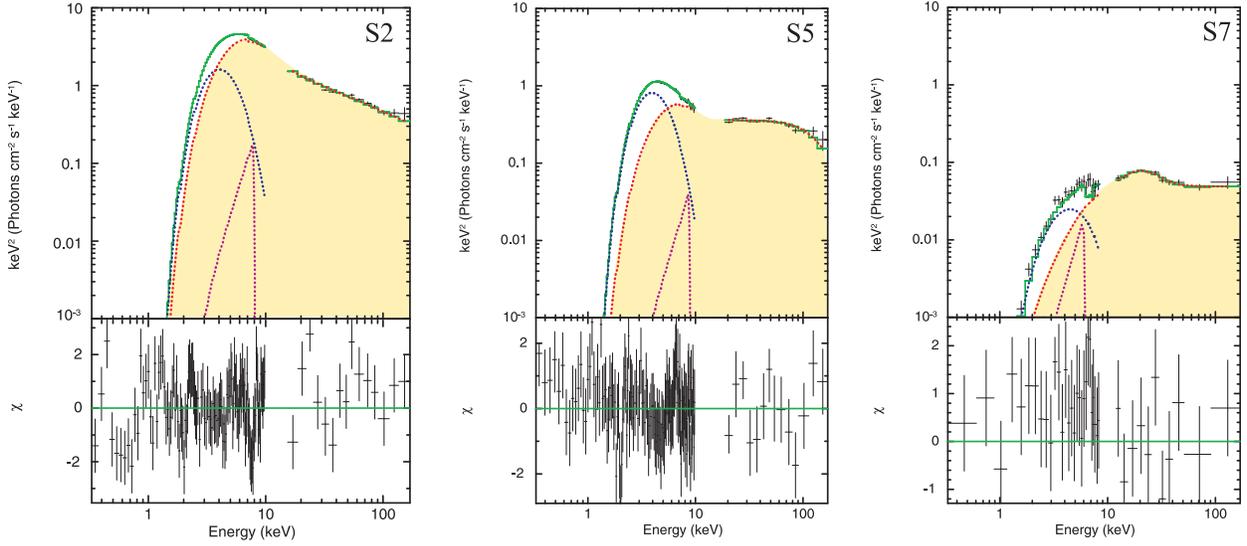}
\caption{Three representative $EF_E$  diagrams for different states of 4U~1630--47 (green lines). 
Data are taken from $Beppo$SAX observations 
20114002 ({\it left} panel, ``S2'' data set, HSS),
20114005 ({\it central} panel, ``S5'' data set, IS),
and 70821005 ({\it right} panel, ``S7'' data set, LHS).
The data are shown by black crosses and  
 the spectral model components are displayed  by dashed red, blue and  purple lines for $Comptb$, 
 $Blackbody$ and $Laor$ respectively. Yellow shaded areas demonstrate an evolution of $Comptb$ component 
during  transitions 
between the HSS ($S2$) and LHS ($S7$)  when the normalization parameter $N_{com}$ of the Comptonization component monotonically decreases from 
13 to 0.5$\times L_{37}/d_{10}^2$ erg/s/kpc$^2$ (see also Fig.~\ref{1655_scal}).
}
\label{BeppoSAX_spectra}
\end{figure}

%
%

\newpage
\begin{figure}[ptbptbptb]
\includegraphics[scale=1.2, angle=0]{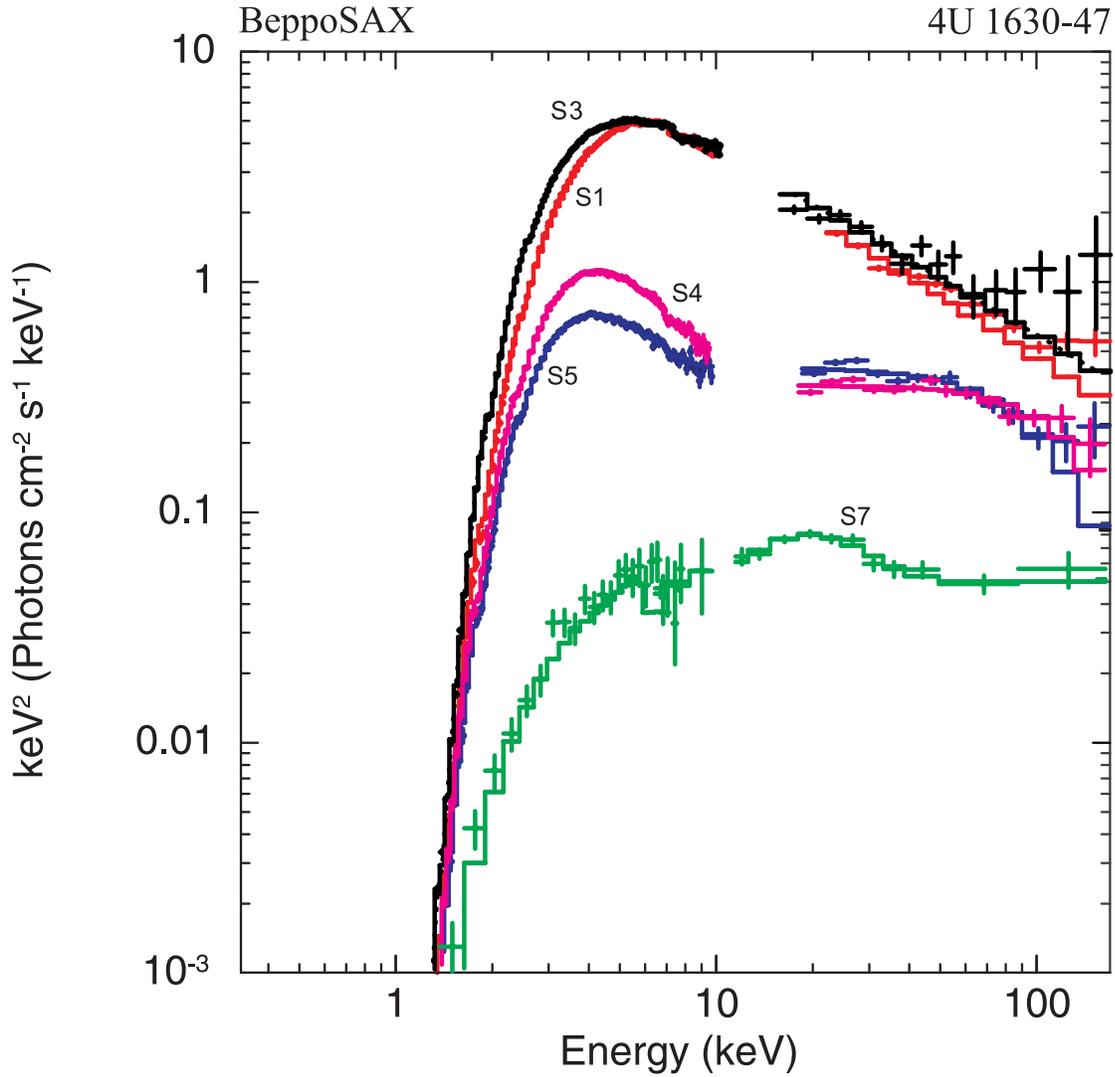}
\caption{Five representative $EF_E$ spectral diagrams for different states of 4U~1630--47. 
Data are taken from $Beppo$SAX observations 
20114001 ($red$), 20114003 ($black$) [HSS], 20114004 ($pink$), 20114005 ($blue$) [IS] and 70821005  ($green$, LHS).
}
\label{sp_compar_SAX}
\end{figure}

%
%





%
%

\newpage
\begin{figure}[ptbptbptb]
\includegraphics[scale=0.9, angle=0]{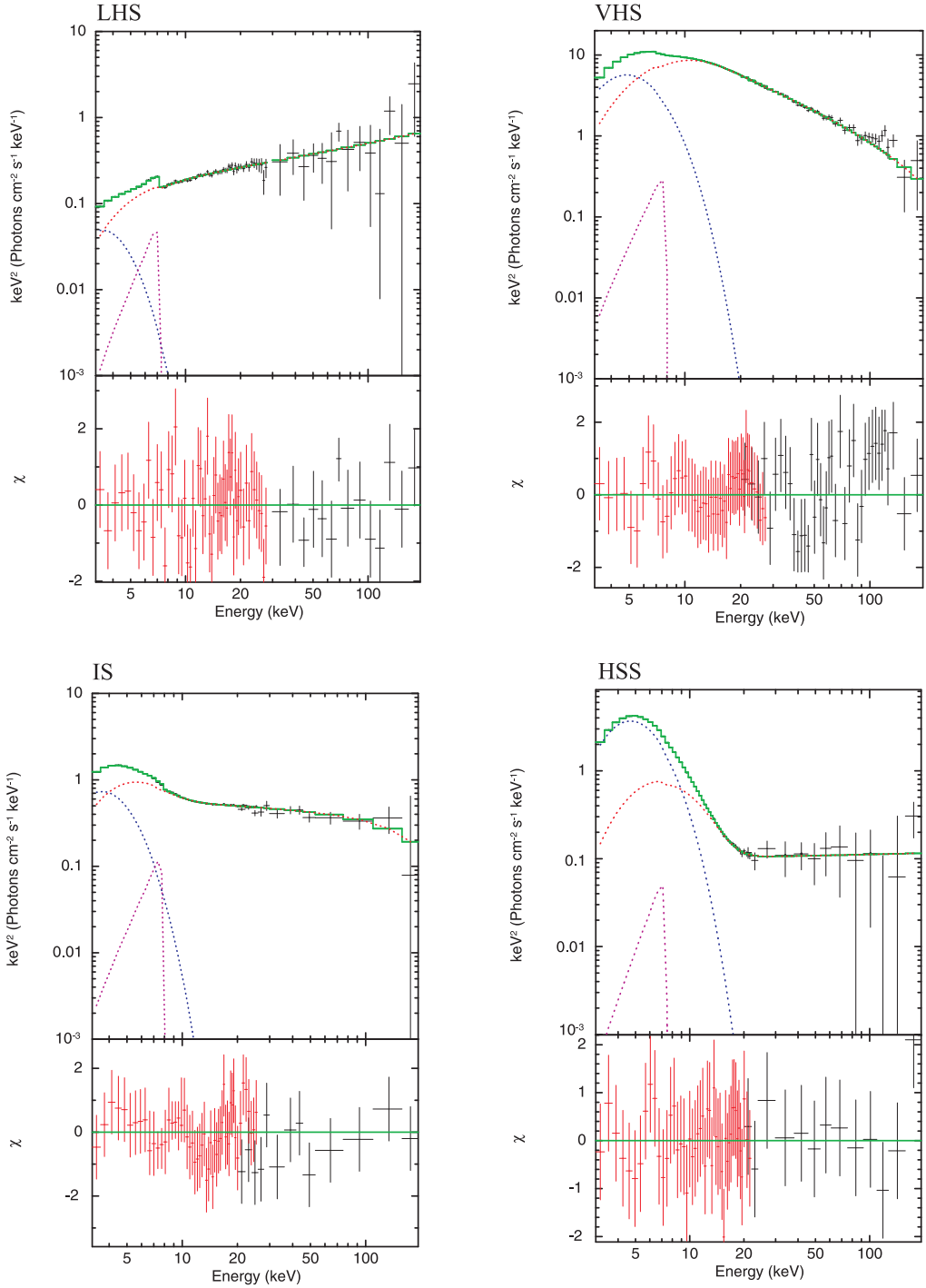}
\caption{
Evolution of spectral diagrams of 4U~1630--47.
Data are taken from {\it RXTE} 
observations 30172-01-18-12 ($\Gamma=1.6$, LHS), 30172-01-04-00 ($\Gamma=2.2$, IS), 80117-01-05-00 ($\Gamma=3.0$, VHS), 
and 10411-01-03-00 ($\Gamma=2.0$, HSS). Here data
are denoted by black points; the spectral model presented with components is shown by blue, red, and purple lines for 
$Blackbody$, $Comptb$, and $Laor$ components, respectively.
}
\label{sp_rxte_evol}
\end{figure}

%
%

\newpage
\begin{figure}[ptbptbptb]
\includegraphics[scale=1.0, angle=0]{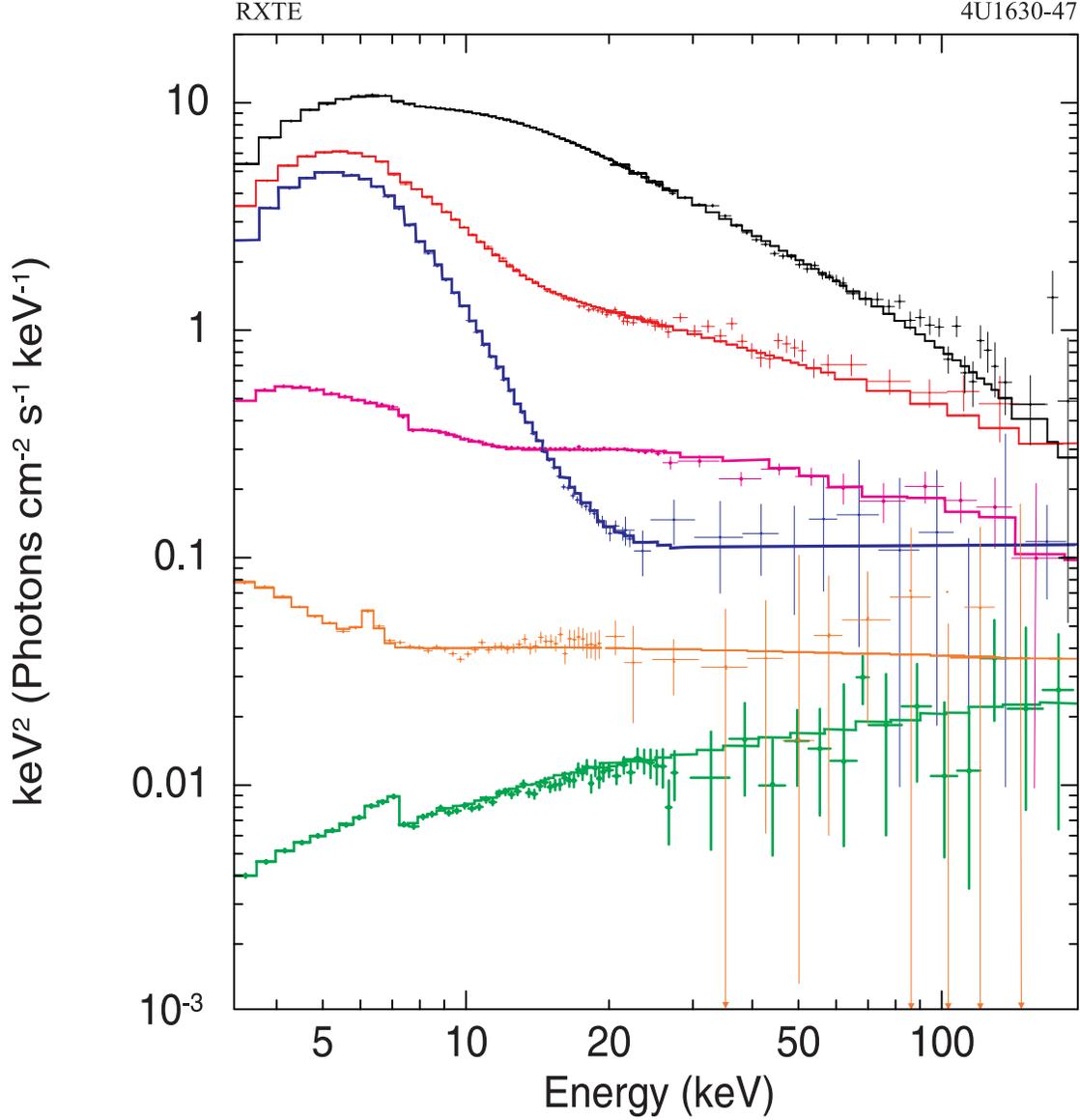}
\caption{Six representative $EF_E$ spectral diagrams which are related to different spectral states 
of 4U~1630--47 using the model $wabs*(Blackbody+Comptb+Laor)$. The  data are taken from {\it RXTE} 
observations 30172-01-18-12 ($green$, LHS), 10411-01-18-00 ($orange$, LHS), 30172-01-07-00 ($pink$, IS), 
70417-01-03-00 ($red$, IS), 10411-01-03-00 ($blue$, HSS), 80117-01-03-00G ($black$, VSS).
The normalization factors of 0.5 and 0.1 was applied for 
10411-01-18-00 and 30172-01-18-12 spectra for clarity.
}
\label{sp_rxte_compar}
\end{figure}

%
%

\newpage
\begin{figure}[ptbptbptb]
\includegraphics[scale=1, angle=0]{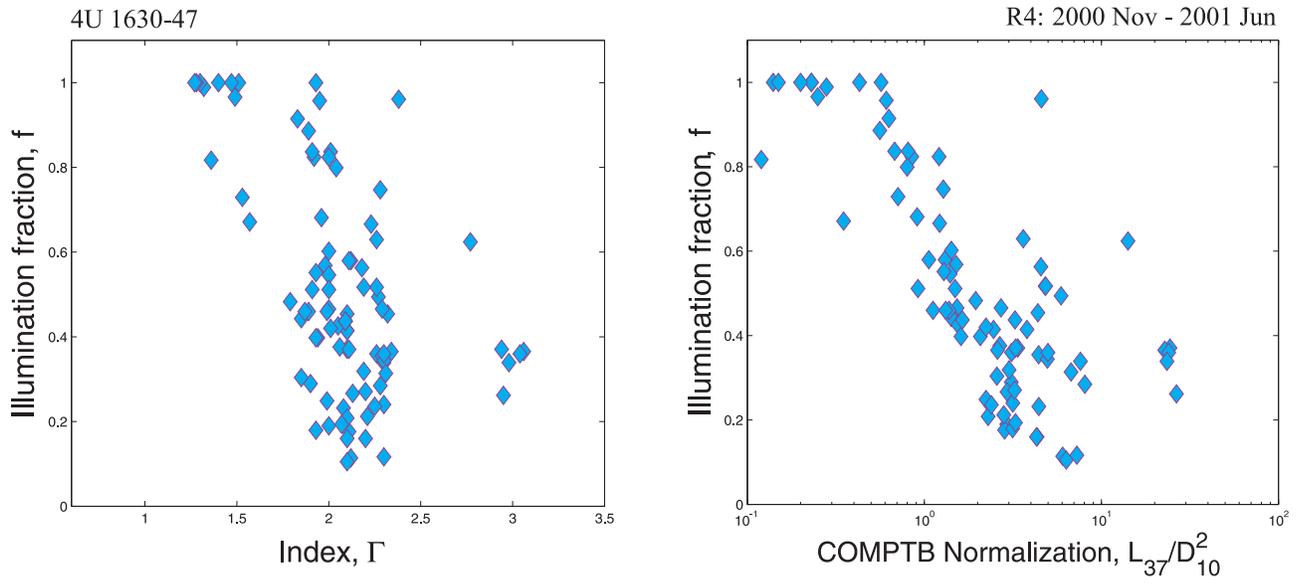}
\caption{
An example of the illumination fraction f vs the photon index $\Gamma$ (left ) and COMPTB normalization (right) for  an observational set
$R4$ (2000 Nov -- 2001 Jun.). }
\label{saturation}
\end{figure}

%
%


%
%

\newpage
\begin{figure}[ptbptbptb]
\includegraphics[scale=1.11, angle=0]{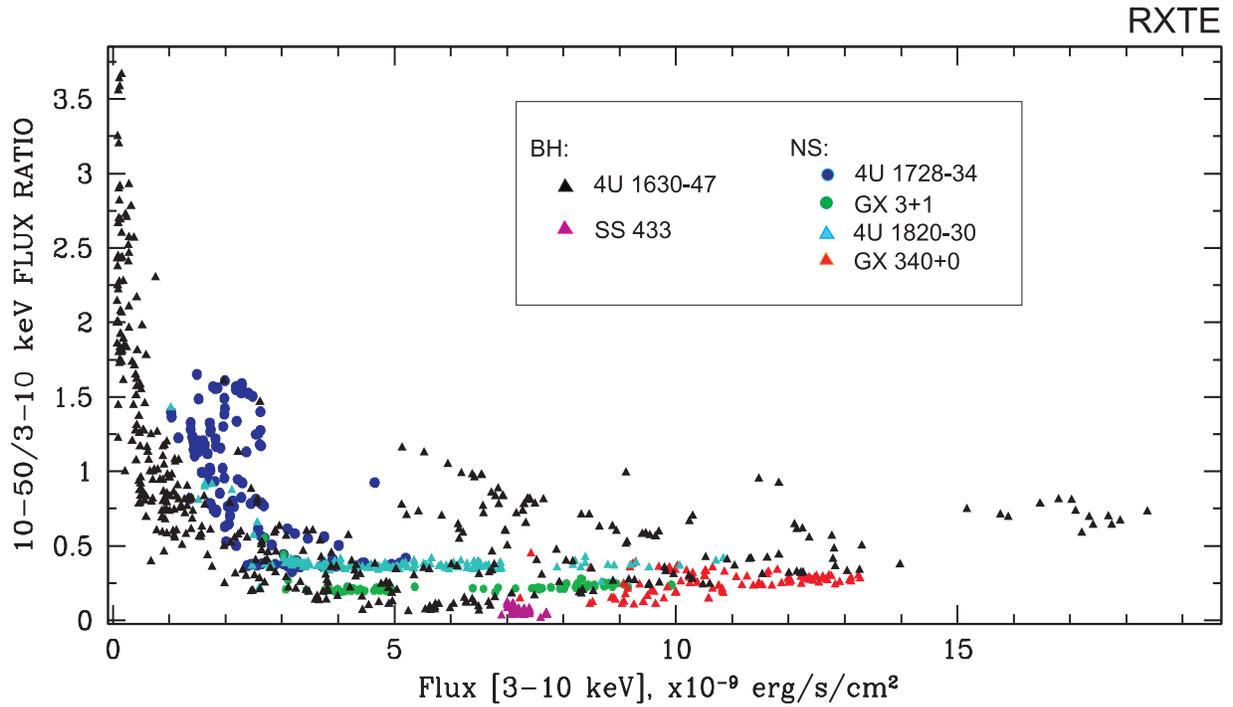}
\caption{
Spectral hardness (10 -- 50 keV/3 -- 10 keV) vs flux in  3 -- 10 keV range of BHCs: 4U~1630--47 ($black$ triangles), 
SS~433 ($violet$ triangles, taken from ST10), and NSs: $atoll$ sources 4U~1728-34 ($blue$ circles, taken from Seifina \& Titarchuk 2011), 
GX~3+1 ($green$ circles, taken from ST12) and 4U~1820-30 ({\it bright blue} triangles), and 
Z source GX~340+0 ($red$ triangles) for {\it RXTE} data.
}
\label{HID_6obj}
\end{figure}

\newpage
\begin{figure}[ptbptbptb]
\includegraphics[scale=0.8,angle=0]{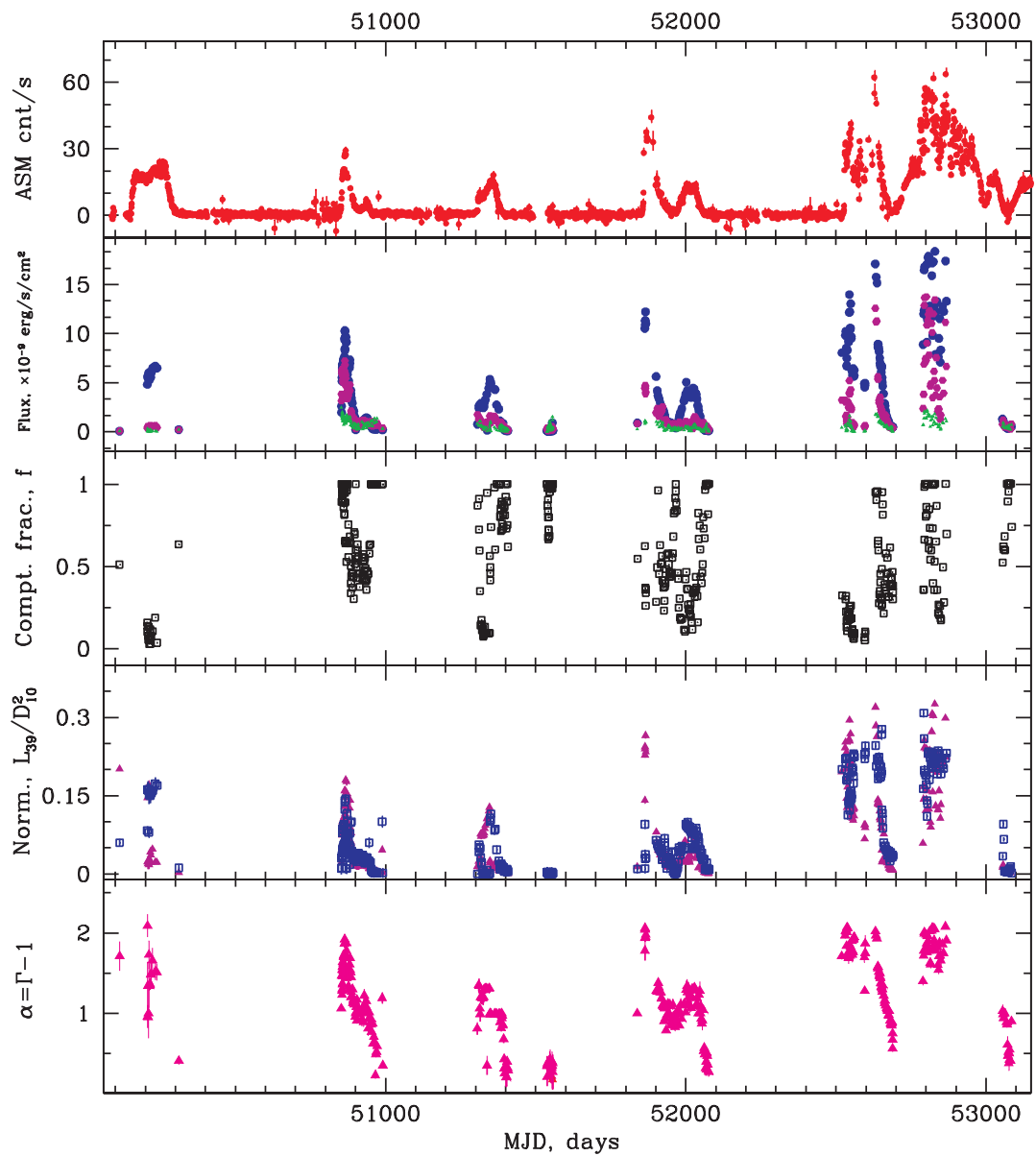}
\caption{
{\it From Top to Bottom:}
Evolutions of the {\it RXTE}/ASM count rate, the model flux in 3-10 keV, 10-50 keV and 50-200 keV energy ranges 
({\it blue, crimson and green} points respectively), the illumination  fraction $f$ 
 and $Comptb$ and $blackbody$ 
normalizations ({\it crimson} and {\it blue} 
respectively)   for all the {\it RXTE} sets ($R1$ -- $R7$, 1996 -- 2004) 
}
\label{evolution_all}
\end{figure}

%
%

\newpage
\begin{figure}[ptbptbptb]
\includegraphics[scale=1.1,angle=0]{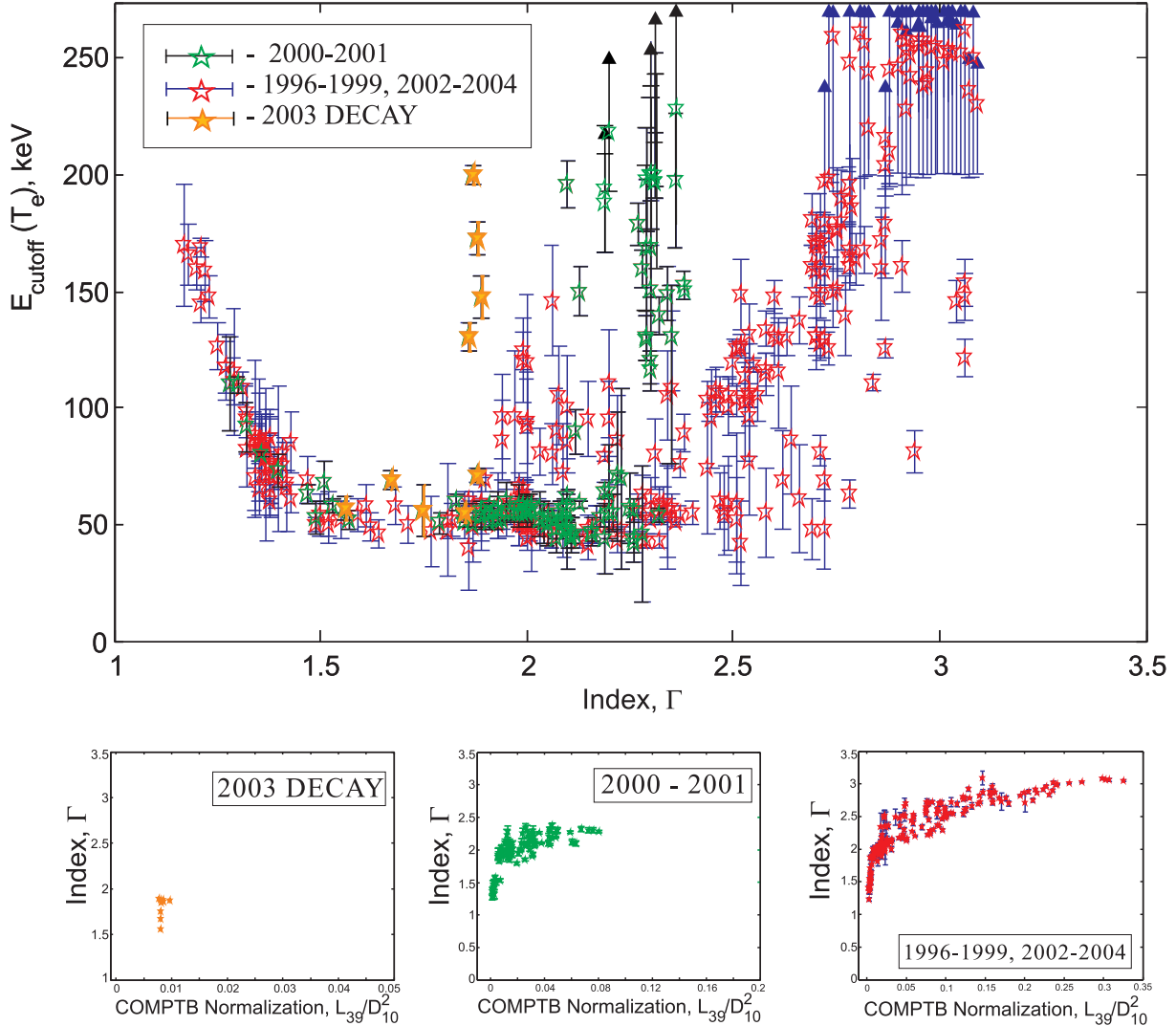}
\caption{
{$Top:$} Cut-off energy $E_{cutoff}$ versus photon index $\Gamma$ for 
{\it RXTE} observations of 4U~1630--47. Spectral parameters 
from 1996-1999 \& 2002-2004/2000-2001/2003[decay] outburst sets are marked in $red$/$green$/$yellow$ stars to 
indicate the saturation levels about $\Gamma\sim 3.0/2.3/1.9$ respectively. 
{$Bottom:$}
$\Gamma$ versus $Comptb$ normalization correlations for all indicated observations: 
decay 2003 ({\it left}), 2000-2001 ({\it center}) and 1996-1999 \& 2002-2004 ({\it right}) outburst sets. 
 Correlation show three different tracks, which correspond to different index
saturation levels. Index saturation level probably correlates with  high
energy cut-off  $E_{cutoff}$(see $top$ panel and the text).
}
\label{gam_t_e}
\end{figure}

%
%

\newpage
\begin{figure}[ptbptbptb]
\includegraphics[scale=0.95,angle=0]{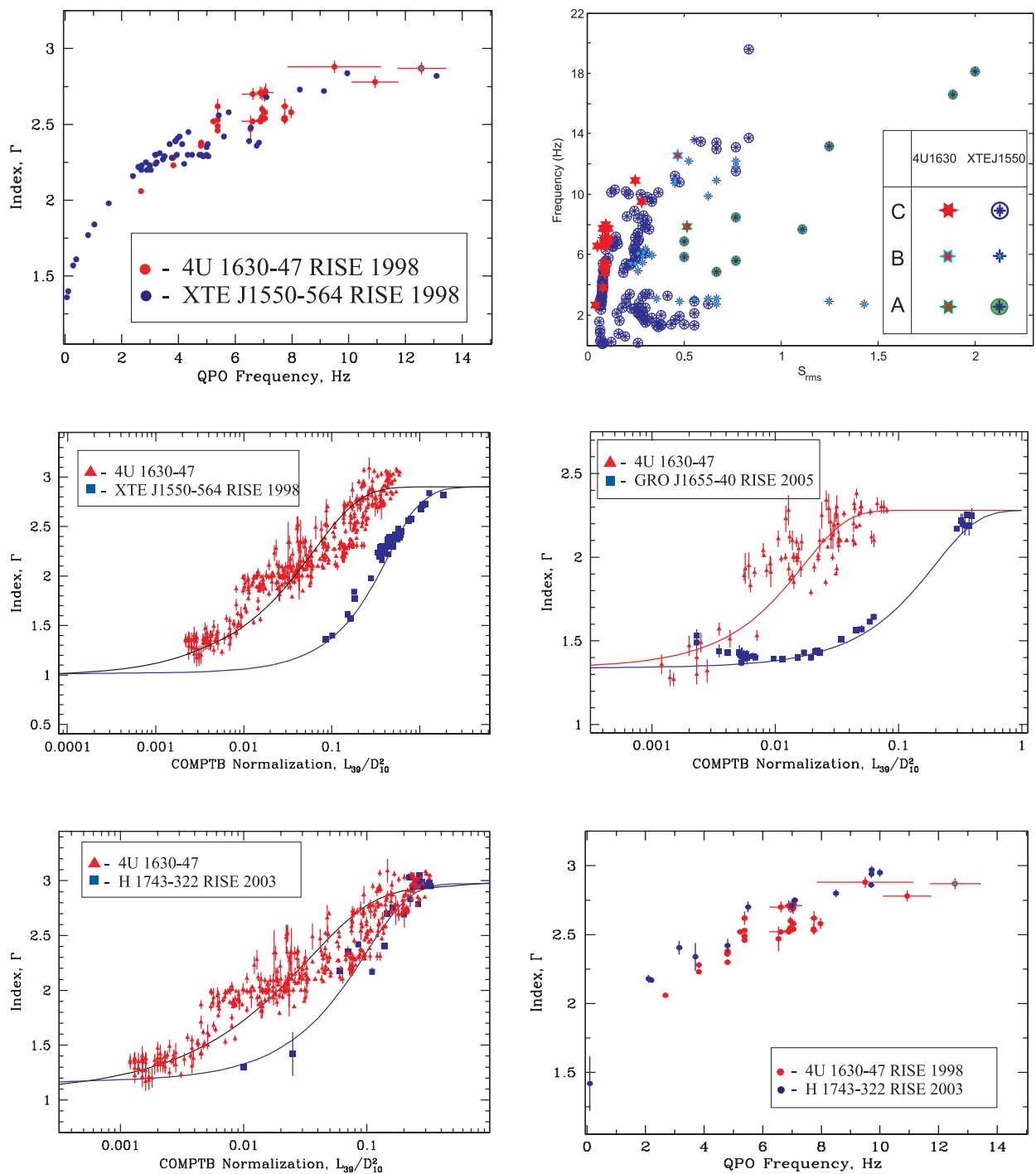}
\caption{Scaling of photon index $\Gamma$ 
for 4U~1630--47 (with $red$ points 
-- target source)  
and XTE J1550-564, GRO~J1655-40 and H1743-322 (with $blue$ marks 
-- reference source). QPO frequency as a function of $S_{rms}$ (the inverse of the rms) for two objects 
(4U~1630-47 and XTE~J1550-564) is presented in {\it right top} panel with the indications of the type A, B and C QPOs.
}
\label{1655_scal}
\end{figure}

%
%

\newpage
\begin{figure}[ptbptbptb]
\includegraphics[scale=0.9, angle=0]{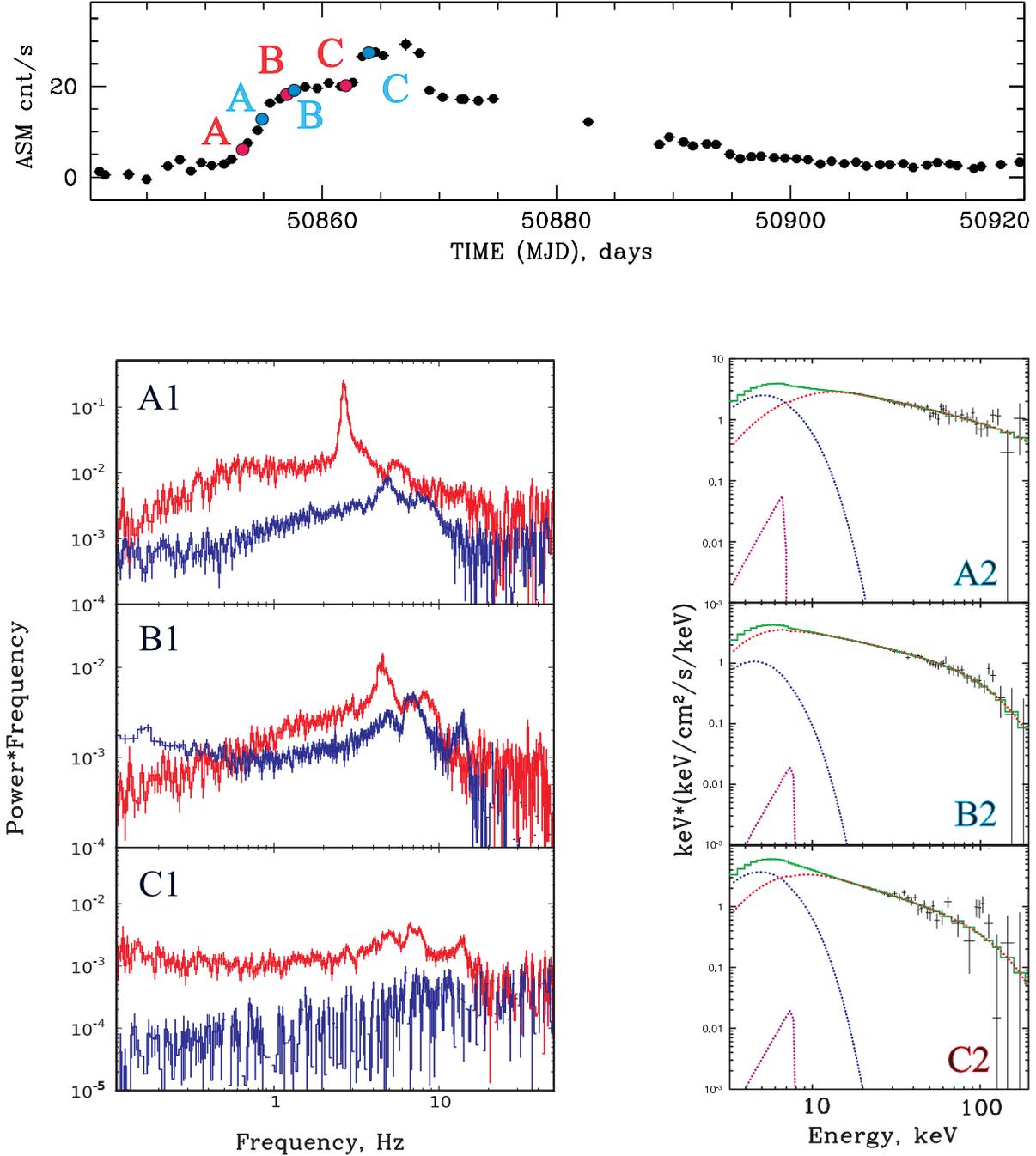}
\caption{
{\it Top}: evolution of the 1.3 -- 12 keV ASM flux during the 1998 rise transition events ($R2$). 
Red/blue points A, B, and C mark moments at 
MJD = 50853.1/50855.8, 50856.1/50857.8 and 50862.6/50864.2 respectively. 
$Bottom$: 
PDSs for 3-13 keV  band ($left$ column) are plotted along with the $E*F(E)$ diagram ($right$ column) 
for A, B and C points  of X-ray light curve. 
$E*F(E)$ diagrams (panels A2, B2, C2) are related to the corresponding power spectra for panels: 
A1 (point A $blue$), B1 (point B $blue$), C1 (point C $red$).  
The data are shown by black points and  
 the spectral model components are displayed  by $red$, $blue$ and $pink$ 
dashed lines for $Comptb$, $Blackbody$ and $Laor$ components  respectively.
}
\label{pds_sp_1998}
\end{figure}

%
%

\newpage
\begin{figure}[ptbptbptb]
\includegraphics[scale=.94,angle=0]{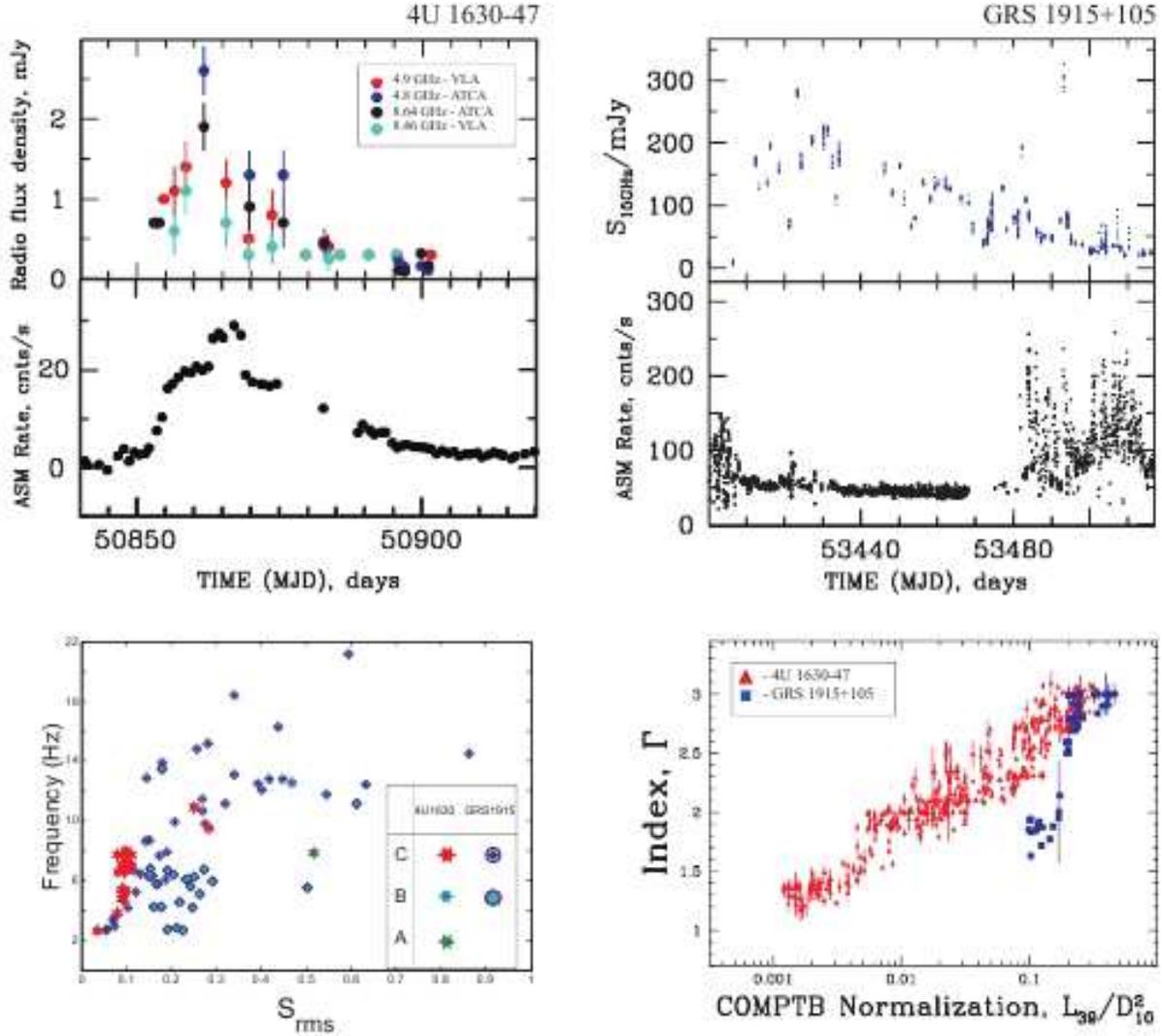}
\caption{
$Top:$ Evolution of flux density 
at 4.9 GHz (VLA, $red$), 4.8 GHz (ATCA, $blue$), 8.64 GHz (ATCA, $blue$) and 8.46 GHz (VLA, $bright~blue$, 
taken from Hjellming et al., 1999) and 
RXTE/ASM count rate during 1998 outburst transition of 4U~1630-47 ($left$ column) 
and 
evolution of flux density $S_{15GHz}$ at 15 GHz (Ryle Telescope, taken from TS09) and RXTE/ASM count rate
during 2005 outburst transition ($right$ column) of GRS 1915+105. 
$Bottom:$ QPO frequency as a function of $S_{rms}$ (the inverse of the rms) for 
4U~1630-47 and GRS~1915+105 (taken from Soleri et al., 2008) is presented 
with the indications of the type-A, -B and -C QPOs ($left$). $Right:$ Photon index $\Gamma$ vs normalization 
for 4U~1630--47 ($red$ points) and GRS~1915+105 ($blue$ points for the 
1997 outburst transition,  taken from TS09).
}
\label{gam_qpo_freq_scal}
\end{figure}



%
%


%
%


\end{document}